\documentclass[12pt]{article}

\usepackage{amsmath}
\usepackage{amssymb}
\usepackage{epsfig}
\usepackage{graphicx}
\usepackage{cite}
\usepackage{multirow}
\usepackage{longtable}
\usepackage{lscape}
\usepackage{bbm}
\usepackage{dcolumn}
\usepackage[dvips]{color}
\usepackage{bm}
\usepackage{booktabs}
\usepackage{wick}
\usepackage{rotating}
\usepackage{mathrsfs}

\allowdisplaybreaks[4] 

\newcommand{\capdef}{}
\newcommand{\mycaption}[2][\capdef]{\renewcommand{\capdef}{#2}%
        \caption[#1]{{\footnotesize #2}}}
\makeatletter
\renewcommand{\fnum@table}{\textbf{\tablename~\thetable}}
\renewcommand{\fnum@figure}{\textbf{\figurename~\thefigure}}
\makeatother

\newcounter{myenumi}

\renewcommand{\themyenumi}{\roman{myenumi}}
{\end{list}}

\newlength{\myem}
\newcounter{mysubequation}[equation]

\makeatletter
\renewcommand{\section}{\@startsection{section}{1}{0em}{-\baselineskip}%
{\baselineskip}{\normalfont\large\bfseries}}
\renewcommand{\subsection}%
{\@startsection{subsection}{2}{0em}{-0.7\baselineskip}%
{0.7\baselineskip}{\normalfont\bfseries}}
\makeatother

\newcommand{\bi}{\begin{itemize}}
\newcommand{\ei}{\end{itemize}}

\newcommand{\be}{\begin{equation}}
\newcommand{\ee}{\end{equation}}
\newcommand{\bea}{\begin{eqnarray}}
\newcommand{\eea}{\end{eqnarray}}
\newcommand{\nn}{\nonumber}
\newcommand{\sss}[1]{\scriptscriptstyle{#1}}

\newcommand{\dv}{\partial\hspace{-7pt}\slash}

\newcommand{\D}{D\hspace{-8pt}\slash}

\newcommand{\ie}{{\it i.e.}}

\newcommand{\eg}{{\it e.g.}}

\newcommand{\cf}{{\it cf.}}

\newcommand{\etc}{{\it etc.}}
\newcommand{\eq}{Eq.}

\newcommand{\fig}{Fig.}

\newcommand{\Ref}{Ref.}
\newcommand{\Refs}{Refs.}
\newcommand{\Sec}{Sec.}

\newcommand{\App}{Appendix}

\newcommand{\Tab}{Table}
\newcommand{\Tabs}{Tables}

\newcommand{\eet}{\epsilon^m_{e\tau}}

\newcommand{\ett}{\epsilon^m_{\tau\tau}}

\newcommand{\emta}{|\epsilon^m_{\mu\tau}|}

\newcommand{\equ}[1]{\eq~(\ref{equ:#1})}
\newcommand{\figu}[1]{\fig~\ref{fig:#1}}

\newcommand{\Slash}[1]{\ooalign{\hfil$ \diagup $\hfil\crcr$#1$}}

\newcommand{\onsi}{\mathcal{O}_{\mathrm{NSI}}}

\begin{document}

\begin{titlepage}

\renewcommand{\thefootnote}{\alph{footnote}}

\vspace*{-3.cm}
\begin{flushright}
\flushright{FTUAM-08-15\\ IFT-UAM/CSIC-08-53\\LPT-ORSAY 08-75
}
\vskip 2.0cm

\end{flushright}

\renewcommand{\thefootnote}{\fnsymbol{footnote}}
\setcounter{footnote}{-1}

{\begin{center}
{\large\bf
Large gauge invariant non-standard neutrino interactions
} \end{center}}
\renewcommand{\thefootnote}{\alph{footnote}}

\vspace*{.8cm}
\vspace*{.3cm}
{\begin{center} {\large{\sc
 		M.~B.~Gavela\footnote[1]{\makebox[1.cm]{Email:}
                belen.gavela@uam.es},
 		D.~Hernandez\footnote[2]{\makebox[1.cm]{Email:}
                hernandez@delta.ft.uam.es},
 		T.~Ota\footnote[3]{\makebox[1.cm]{Email:}
                toshihiko.ota@physik.uni-wuerzburg.de}, and
                W.~Winter\footnote[4]{\makebox[1.cm]{Email:}
                winter@physik.uni-wuerzburg.de}
                }}
\end{center}}
\vspace*{0cm}
{\it
\begin{center}

\footnotemark[1]${}^,$\footnotemark[2]
	Departamento de F{\'i}sica Te{\'o}rica and Instituto de F{\'i}sica Te{\'o}rica UAM/CSIC, \\
        Universidad Aut{\'o}noma de Madrid, 28049 Cantoblanco, Madrid, Spain 

\footnotemark[2]
	Laboratoire de Physique Th\'{e}orique, UMR CNRS 8627
	bat. 210, Universit\'{e} Paris-Sud X1, 91405, Orsay, France
	
\footnotemark[3]${}^,$\footnotemark[4]
       Institut f{\"u}r Theoretische Physik und Astrophysik, \\ Universit{\"a}t W{\"u}rzburg, 
       97074 W{\"u}rzburg, Germany

\end{center}}

\vspace*{1.5cm}

{\Large \bf
\begin{center} Abstract \end{center}  }

Theories beyond the Standard Model must necessarily respect its gauge symmetry. 
This implies strict constraints on the possible models of non-standard neutrino interactions, which we analyze. The focus is set on the effective low-energy dimension six and eight operators  involving four leptons, decomposing them according to all possible tree-level mediators, as a guide for model building.
 The new couplings are required to have sizeable strength, while processes involving four charged leptons are required to be suppressed. 
 For non-standard interactions in matter, only diagonal tau-neutrino interactions can escape these requirements and can be allowed to result from dimension six operators.  
 Large non-standard neutrino interactions from dimension eight operators alone are phenomenologically allowed in all flavour channels and shown to require at least two new mediator particles.  The new couplings must obey general cancellation conditions both at the dimension six and eight levels, which result from expressing the operators obtained from the mediator analysis in terms of a complete basis of operators.
  We illustrate with one example how to apply this information to model building.

\end{titlepage}

\newpage

\renewcommand{\thefootnote}{\arabic{footnote}}
\setcounter{footnote}{0}

\section{Introduction}

The experimental observation of neutrino masses and mixings is the first evidence for physics beyond the Standard Model (SM) -- maybe together with the indication for dark matter -- and points to the existence of a new, yet unknown, physics scale.  The tiny masses of the neutrinos, which are orders of magnitude lighter than those of other fermions, suggest a large new physics scale leading to very suppressed effects. 
Since neutrinos have only weak interactions with the SM particles, they may even
constitute an excellent window into the new physics underlying the
``dark sectors'' of the universe, \ie, dark matter and dark energy. 
Therefore,  new physics may very well appear next in the form of exotic couplings involving neutrinos, which are often called {\it non-standard neutrino interactions} (NSI)~\cite{Wolfenstein:1977ue,Valle:1987gv,Guzzo:1991hi,Grossman:1995wx,Roulet:1991sm}. 
This possibility is being actively explored and will be the subject of the present work. 
In particular, we will, in a model-independent way, discuss the connection between NSI and the possible tree-level mediators of new physics inducing them. If NSI are detected, this study will serve as a guide for the model builder. 

Note that the very tiny neutrino mass differences have only been
 detectable because the masses affect the neutrino propagation by
 inducing small phase shifts, which can be compensated by the very long
 distances travelled in neutrino oscillation experiments. In contrast,
 non-standard couplings are short distance (local) effects,  which
 usually do not benefit from such an enhancement -- unless they affect
 the propagation in matter, one of the several possibilities explored
 below. Notice, though, that neutrino oscillation experiments may well turn out to be
  the best {\it arena} to detect NSI, as they can affect oscillation amplitudes linearly, 
  instead of quadratically as in most charged lepton flavour violation processes.

 On general grounds, whatever the nature of the new putative couplings is, observable effects will only be expected for a new physics scale $\Lambda$ near the present experimental limits, \ie, above the electroweak symmetry breaking (EWSB) scale. 
An example of NSI is given by  the dimension six ($d=6$) operator in 
\begin{equation}
\frac{1}{\Lambda^2} (\bar{\nu}_\alpha \gamma^\rho P_L \nu_\beta)\,( \bar\ell_\gamma \gamma_\rho \ell_\delta )\, .
\label{equ:raw}
\end{equation}
In this expression, spinor indices are omitted, Greek letters denote flavour indices, $P_L=(1 - \gamma_5)/2$ is the left-handed chiral projection operator, and $\nu$ and $\ell$  label the SM neutrino and charged lepton fields, respectively.

The operator in \equ{raw} is not a singlet of the SM gauge group, $SU(3)\times SU(2)\times U(1)$, while the high-energy theory has to contain and encompass the SM gauge group, though. 
For instance, the coupling in \equ{raw} could result from the following gauge invariant operator:
\begin{equation}
\frac{1}{\Lambda^2} (\bar{L}_\alpha \gamma^\rho L_\beta)\,( \bar{L}_\gamma \gamma_\rho L_\delta \,) \, ,
\label{equ:dressed}
\end{equation}
where $L$ denotes the leptonic $SU(2)$ doublets. 
\equ{dressed} illustrates the consequences of electroweak gauge
invariance: The coupling in \equ{raw} comes necessarily accompanied by
other charged lepton transitions  for which stringent limits  may
exist. As an example, for $\beta=\mu$ and $\alpha=\gamma=\delta=e$,
\equ{dressed} would imply  $\mu\rightarrow 3\, e$ transitions with the
same strength than the interaction in \equ{raw}.

 In order to discuss such constraints coming from the gauge invariant framework, it is convenient to rephrase the problem in terms of a generic  low-energy effective theory.  Effective theories allow rather model-independent analyses based on the fundamental symmetries, while only the coefficients of the effective operators are  model-dependent.
  The impact of the heavy fields present  in  the high-energy theory  
 can be parametrized, without loss of generality, by the addition to the Lagrangian of a tower  of {\it non-renormalizable} 
 operators $\mathcal{O}^{d}$ of dimension $d>4$, made out of the SM fields and invariant under the SM gauge group. The operator coefficients are weighted by inverse powers of the high scale $\Lambda$: 
 \be\label{leff}
\mathscr{L} = \mathscr{L}_{\rm SM} + \delta\mathscr{L}^{d=5}_{\text{eff}} 
+ \delta\mathscr{L}^{d=6}_{\text{eff}} + \cdots
\, , \quad \textrm{with} \quad \delta\mathscr{L}^{d}_{\text{eff}} \propto \frac{1}{\Lambda^{d-4}} \mathcal{O}^{d}\,.
\ee
After EWSB, some of the effective operators may result in corrections to the low-energy SM parameters. In addition, new exotic couplings may result, such as those in Eq.~(\ref{equ:raw}).

The only possible $d=5$ operator   is the famous Weinberg operator~\cite{Weinberg:1979sa}, which leads, after EWSB, to Majorana neutrino masses.
  We will not need to consider it for the present study of NSI. 

At $d=6$, some operators  modify the low-energy standard couplings, among which non-unitary corrections to the leptonic
  PMNS mixing matrix are specially relevant to the case under study. Departures from unitarity are a general feature of
  beyond the SM models involving exotic
  fermions~\cite{Valle:1987gv,Langacker:1988up,Langacker:1988ur,Broncano:2002rw,Broncano:2003fq,Abada:2007ux,Czakon:2001em,Bekman:2002zk,Holeczek:2007kk,delAguila:2008pw}.
  All fermions of the same charge will generically mix through the mass
  matrix, leading in those theories to unitary mixing matrices of dimension larger than three,
  while the effective $3\times 3$ sub-matrix  - relevant at low energies - is
  not unitary. In the effective Lagrangian formalism, the effect appears technically at leading order through $d=6$ gauge invariant operators involving only {\it two} fermions, which induce non-canonical corrections to the fermion kinetic terms~\cite{Langacker:1988up,Langacker:1988ur,Broncano:2002rw,Broncano:2003fq,Abada:2007ux}. Such operators are for instance typical of fermion-mediated Seesaw scenarios.
  The trademark of non-unitarity is that the coefficients of the NSI
  operators induced by it and contributing to neutrino production,
  detection, and matter effects are not independent but related. Barring
  fine-tuned cancellations,  the stringent bounds and future signals on
  non-unitarity~\cite{Farzan:2002ct,Sato:2000wv,Barger:2004db} apply as
  well to NSI, except for those NSI operators affecting exclusively the
  propagation in matter.  Recently,  the value of the elements of the
  PMNS matrix have been extracted from data without assuming a unitary mixing
  matrix~\cite{Antusch:2006vwa}, and new related CP-odd signals have
  been proposed as well~\cite{FernandezMartinez:2007ms,Goswami:2008mi,Altarelli:2008yr}. For a detailed discussion of the NSI-non-unitarity relationship, see~\cite{Antusch:2008tz}. We will leave non-unitarity out of the main line of this work, discussing only its qualitative implications.

Effective interactions such as the NSI in \equ{raw}, obviously require to consider operators made out of {\it four} leptonic fields, plus Higgs fields in the case of operators with $d> 6$~\cite{Davidson:2003ha,Bergmann:1998ft,Bergmann:1999pk}.
There is a plethora of $d=6$~\cite{Buchmuller:1985jz} and $d=8$~\cite{Berezhiani:2001rs} operators~\footnote{ $d=7,9...$ operators~\cite{deGouvea:2007xp} are odd under baryon and lepton number and not relevant for the present discussion.}
, with different classes of models resulting in different operators and operator coefficients. 
 Those among them relevant for NSI can affect neutrino production or detection processes, or modify the matter effects in their propagation, depending on the operator or combination of operators considered.  Notice, though, that the coefficient of a $d=8$ operator is expected to be suppressed by a factor $~v^2/\Lambda^2$ with respect to $d=6$ operator coefficients (where $v$ is the vacuum expectation value of the Higgs field $\langle
H^{0} \rangle = v/\sqrt{2} = 174$ GeV), and thus negligible unless the new scale is very close to the electroweak one.

Much effort has been dedicated to analyze the experimental constraints
  and future detection prospects of
  NSI~\cite{Davidson:2003ha,Barranco:2007ej,Yao:2006px}. This
  encompasses their impact on weak decays, solar and atmospheric
  neutrino physics~\cite{Gonzalez-Garcia:1998hj,Bergmann:2000gp,Fornengo:2001pm,Miranda:2004nb,Friedland:2004ah,Gonzalez-Garcia:2004wg,Friedland:2005vy,GonzalezGarcia:2007ib}, 
  astrophysics~\cite{Fogli:2002xj,EstebanPretel:2007yu}, early universe processes~\cite{Mangano:2006ar}, collider and neutrino scattering experiments~\cite{Berezhiani:1994hy,Berezhiani:2001rs,Berezhiani:2001rt,Chen:2007cn, Barranco:2007tz},   
  and past and future neutrino oscillation experiments including neutrino factories~\cite{Huber:2001zw,Gonzalez-Garcia:2001mp,Ota:2001pw,Gago:2001xg,Campanelli:2002cc,Ota:2002na,Huber:2002bi,Hattori:2002uw,Garbutt:2003ih,Blennow:2005qj,Friedland:2006pi,Kitazawa:2006iq,Honda:2006gv,Blennow:2007pu,Kopp:2007mi,Ribeiro:2007ud,Kopp:2007ne,Ribeiro:2007jq,Kopp:2008ds,EstebanPretel:2008qi,Blennow:2008ym,Blennow:2008eb,Winter:2008eg}. Overall, the signals involving the $\nu_\tau$ field are at present the least constrained ones.

We emphasize that we study ``large'' NSI, where ``large'' refers to being potentially observable at future neutrino oscillation experiments. For example, for the flavor-changing NSI interfering with standard oscillations, bounds $10^{-2}$ to $10^{-3}$ (relative to the SM four-fermion interaction coupling) on the operator in \equ{raw} can be expected from a neutrino factory (see, \eg, \Ref~\cite{Kopp:2008ds}). After EWSB, the four-fermion interactions from Eq.~(\ref{leff})
will be suppressed by roughly $(v/\Lambda)^{d-4}$ with respect to the  SM four-fermion interactions,
which means that $\Lambda$ is allowed to be largest for the $d=6$ effective operators, whereas for $d=8$ operators, it has to be very close to the electroweak scale in order to produce a sizable effect. Therefore, we focus on the $d=6$ operators first, and then increase the level of complexity.
However, from this simple comparison, one can already read off that new physics above the TeV scale will be very difficult to be observed at future neutrino oscillation experiments, since the suppression with respect to the SM is roughly $10^{-2}$ (for $d=6$) and $10^{-4}$ (for $d=8$), respectively -- especially if the $d=6$ operators turn out to be not good enough.

Most of the literature deals mainly with NSI in experiments, and does
not discuss the models behind them -- with some
exceptions~\cite{Guzzo:1991hi,Roulet:1991sm,Bergmann:1998ft,Bergmann:1999pk,Bergmann:1999rz,Ota:2005et,Adhikari:2006uj,Honda:2007wv,Antusch:2008tz}.
In this work, we will classify all $d=6$ and $d=8$, $SU(3)\times
SU(2)\times U(1)$ invariant, leptonic NSI operators in terms of the heavy
mediators inducing them. This can be achieved in a model independent way
decomposing each operator into all possible products of currents. 
The SM charges of the corresponding  field combinations will be the SM charges of the putative heavy mediators. The study will thus be confined to the case in which the tower of non-renormalizable operators has been produced by the {\it tree-level} exchange of new heavy fields, whose precise nature other than their SM charges we do not need to know.

We will first emphasize  the case in which the mediators exchanged only couple to SM bilinear field combinations.  
  In this study, we refer to ``SM bilinears'' as fundamental interactions
of {\it exactly two} SM fields with one or two exotic fields, where the latter
possibility amounts to couplings between two exotic bosons and two Higgs doublets. Other than that, 
there can be in addition new exotic couplings involving only {\it one} SM field, which will also be addressed later on.

The decomposition in terms of SM bilinears, initiated in~\cite{Bergmann:1998ft,Bergmann:1999pk}, immediately leads to correlations between previously uncorrelated effective NSI operator coefficients, with very fruitful physics consequences, as we will show below. 
For instance, it has been realized that the lowest dimension operators, 
that lead to NSI without simultaneously inducing dangerous transitions among four
charged leptons, are $d=8$ operators. The first example proposed~\cite{Davidson:2003ha,Ibarra:2004pe}
is of the form
\be \label{equ:nsi}
\onsi = (\bar{L}^{i} H_{i})\gamma^\rho(H^{\dagger i} L_{i})(\bar{E}\gamma_\rho E) \, ,
\ee
where $i$, $L$ $E$ and $H$ denote  the  $SU(2)_{L}$ index, leptonic doublet, leptonic singlet and Higgs doublet, respectively (note that we use the convention with $Y=-1/2$ for the Higgs doublet, $H^T = (\frac{1}{\sqrt{2}}H^0, H^-)$).
We will illustrate below, though, that any realistic model responsible
for it will also induce other dangerous $d=6$ operators and/or some
other low-energy effects for which stringent experimental bounds may
exist ({\ie} non-unitary corrections to the PMNS matrix, corrections to the 
electroweak precision data, 
flavour-changing neutral currents (FCNC), etc.). 
\equ{nsi} is only one example of the NSI seeked. We will determine in
this work several other independent $d=8$ operators which lead to large
NSI and no charged lepton processes. The symbol $\onsi$ will be extended to denote generically any element in this ensemble.

 In our general analysis, after determining all possible mediators, the resulting correlations between the possible  $d=6$ and $d=8$ operators  will be systematically studied.  We will then establish which mediators or combinations of mediators can lead to large NSI, without inducing experimentally excluded leptonic charged flavour-changing transitions, and/or other undesired phenomenological consequences.

Our main motivation in this study is to determine the minimum level of complexity needed for a viable model of NSI.
 As an illustration for model building, a particular simple toy model will be
 developed in which the operator $\onsi$ above is induced unaccompanied
 by any leptonic $d=6$ operator. 
The aim is to show the generic prize to pay at the theoretical level for allowing observable NSI effects at future experiments. 

Note that we focus in this study on the necessary conditions to build a model with large NSI, while for any given model 
 additional limitations may arise. Supplementary constraints which could arise from a phenomenological analysis at one-loop are also not considered here and should be addressed when considering a particular model.  From the experimental point of view, we will not make any explicit statement how likely it is to observe large NSI.
 We leave the interpretation of this likeliness by judging the necessary conditions for a viable model to the reader. Finally, possible NSI involving quark fields are neither included in this study. 

 \section{Effective operator formalism}
\label{formalism}

 The SM Lagrangian is extended to accommodate the tower of effective operators 
 \begin{equation}
\mathcal{\delta} \mathscr{L }_{\text{eff}} = \frac{1}{\Lambda^2}\sum^{d=6}_i \mathcal{C}_{i} \, \mathcal{O}_i^{d=6} + \frac{1}{\Lambda^4}\sum^{d=8}_k \mathcal{C}_{k} \, \mathcal{O}_k^{d=8}\,,
\label{equ:effop}
 \end{equation}
 where the two terms run over all possible $d=6$ and $d=8$ operators
 relevant for purely leptonic NSI. The flavour composition will be made
 explicit in each coefficient and operator, \ie,
 $(\mathcal{C}_{i})^{\alpha\gamma}_{\beta\delta} (\mathcal{O}_{i})_{\alpha\gamma}^{\beta\delta}$. All distinct flavour combinations for the same operator structure will be taken into account, as they correspond in fact to independent operators.

\subsubsection*{Effective operator basis}

In order to find all possible $d=6$ and $d=8$  effective operators leading to purely leptonic NSI, we will use the following bases:
 \begin{itemize}
 \item {\bf $\boldsymbol{d=6}$ operators}. A complete basis of $d=6$ operators invariant under the SM gauge group and made out of the SM light fields  was proposed by Buchm{\"u}ller and Wyler (BW)~\cite{Buchmuller:1985jz}. The four fermion operator structures relevant to our problem are:
\begin{align}
 (\mathcal{O}_{LE})^{\beta \delta}_{\alpha \gamma}
 =&
(\bar{L}^{\beta} E_{\gamma}) (\bar{E}^{\delta} L_{\alpha})  \, , \label{equ:ole} \\
(\mathcal{O}_{LL}^{{\bf 1}})^{\beta \delta}_{\alpha \gamma}
 =&
 (\bar{L}^{\beta} \gamma^{\rho} L_{\alpha})
 (\bar{L}^{\delta} \gamma_{\rho} L_{\gamma}) \, ,
\label{equ:oll1}
 \\
 (\mathcal{O}_{LL}^{{\bf 3}})^{\beta \delta}_{\alpha \gamma}
 =&
 (\bar{L}^{\beta} \gamma^{\rho} \vec{\tau} L_{\alpha})
 (\bar{L}^{\delta} \gamma_{\rho} \vec{\tau} L_{\gamma}) \, ,
\label{equ:oll3} \\
(\mathcal{O}_{EE})^{\beta \delta}_{\alpha \gamma} =& 
 (\bar{E}^\beta \gamma^{\rho}E_\alpha )(\bar{E}^\delta \gamma_{\rho} E_\gamma)  \, ,
\label{equ:oee}
\end{align}
where $L$ ($E$) refers to the $SU(2)$ leptonic doublet (singlet).
We will refer to the coefficient matrices for these operators by 
       $(\mathcal{C}_{LE})^{\alpha \gamma}_{\beta \delta}$,
       $(\mathcal{C}_{LL}^{{\bf 1}})^{\alpha \gamma}_{\beta \delta}$,
       $(\mathcal{C}_{LL}^{{\bf 3}})^{\alpha \gamma}_{\beta \delta}$,
       and $(\mathcal{C}_{EE})^{\alpha \gamma}_{\beta \delta}$,
       respectively. The operators $\mathcal{O}_{EE}$ do not produce NSI
       directly, but will play a role when considering charged lepton
       flavor violation (since they share some mediators with the operators in Eq.~(\ref{equ:ole})-(\ref{equ:oll3})).

On top of the above, there are two $d=6$ operator structures
including two lepton doublets $L$ and two Higgs doublets $H$,       
\begin{align}
(\mathcal{O}_{LH}^{{\bf 1}})^{\beta}_{\alpha}
=&
\left(
 \bar{L}^{\beta} H
\right)
{\rm i} \dv
(H^{\dagger} L_{\alpha}),\label{odv}
\\
(\mathcal{O}_{LH}^{{\bf 3}})^{\beta}_{\alpha}
=&
\left(
 \bar{L}^{\beta} \vec{\tau} H
\right)
{\rm i} \D
(H^{\dagger} \vec{\tau} L_{\alpha}),\label{oDv}
\end{align}
and a operator with two $E$'s and two $H$'s
\begin{align}
(\mathcal{O}_{EH})^{\beta}_{\alpha}
=&
 \left(
 H^{\dagger}
 {\rm i} D^{\rho}
 H
 \right)
\left(
 \bar{E}^{\beta} \gamma_{\rho} E_{\alpha}
 \right)\,,
 \label{eq:OEH}
\end{align}
where $D$ denotes the SM covariant derivative. These three operators belong to the class which, after EWSB,  correct
       the parameters of the SM Lagrangian. 
       In particular, they renormalize the kinetic energy of neutrinos
       and/or charged
       leptons~\cite{Broncano:2002rw,Broncano:2003fq,Abada:2007ux}. As previously mentioned, 
they result in non-unitary corrections to the the leptonic mixing matrix
       and/or correct the charged and neutral electroweak currents~\cite{Buchmuller:1985jz},  and 
 will not be further developed in this work. We include them above only
       for the sake of completeness (see also Sec.\ref{Sec:Topology1}).
       
 \item {\bf $\boldsymbol{d=8}$ operators}. A basis was discussed by
Berezhiani and Rossi (BR)~\cite{Berezhiani:2001rs}, with the relevant operators given by
\begin{align}
 (\mathcal{O}_{LEH}^{\bf 1})^{\beta \delta}_{\alpha \gamma} 
 =&
 (\bar{L}^{\beta} \gamma^{\rho}  L_{\alpha})
 (\bar{E}^{\delta} \gamma_{\rho}  E_{\gamma})
 \left(H^{\dagger} H\right), \label{equ:br1}
 \\
 (\mathcal{O}_{LEH}^{\bf 3})^{\beta \delta}_{\alpha \gamma} 
 =&
 (\bar{L}^{\beta} \gamma^{\rho} \vec{\tau}  L_{\alpha})
 (\bar{E}^{\delta} \gamma_{\rho}  E_{\gamma})
 \left(H^{\dagger} \vec{\tau} H \right), \label{equ:br2} \\
  (\mathcal{O}_{LLH}^{{\bf 111}})^{\beta \delta}_{\alpha \gamma}
  =&
  (\bar{L}^{\beta} \gamma^{\rho} L_{\alpha})
  (\bar{L}^{\delta} \gamma_{\rho} L_{\gamma})
  (H^{\dagger} H), \label{equ:br3} \\
  (\mathcal{O}_{LLH}^{{\bf 331}})^{\beta \delta}_{\alpha \gamma}
  =&
  (\bar{L}^{\beta} \gamma^{\rho} \vec{\tau} L_{\alpha})
  (\bar{L}^{\delta} \gamma_{\rho} \vec{\tau} L_{\gamma})
  (H^{\dagger} H), \label{equ:br4} \\
  (\mathcal{O}_{LLH}^{{\bf 133}})^{\beta \delta}_{\alpha \gamma}
  =&
  (\bar{L}^{\beta} \gamma^{\rho} L_{\alpha})
  (\bar{L}^{\delta} \gamma_{\rho} \vec{\tau} L_{\gamma})
  (H^{\dagger} \vec{\tau} H), \label{equ:br5} \\
 (\mathcal{O}_{LLH}^{{\bf 313}})^{\beta \delta}_{\alpha \gamma}
 =&
  (\bar{L}^{\beta} \gamma^{\rho} \vec{\tau} L_{\alpha})
  (\bar{L}^{\delta} \gamma_{\rho}  L_{\gamma})
  (H^{\dagger} \vec{\tau} H), \label{equ:br6} \\
  (\mathcal{O}_{LLH}^{{\bf 333}})^{\beta \delta}_{\alpha \gamma}
  =&
  (-{\rm i}\epsilon^{abc})
  (\bar{L}^{\beta} \gamma^{\rho} \tau^{a} L_{\alpha})
  (\bar{L}^{\delta} \gamma_{\rho} \tau^{b} L_{\gamma})
  (H^{\dagger} \tau^{c} H) , \label{equ:br7} \\
 (\mathcal{O}_{EEH})^{\beta \delta}_{\alpha \gamma} =& 
 (\bar{E}\gamma^{\rho}E)(\bar{E} \gamma_{\rho} E)  
 (H^{\dagger} H)   \, . \label{equ:br8}
\end{align}
In these operators, subscripts correspond to a shortcut notation for  their SM field composition,
         whereas superscripts denote the corresponding $SU(2)$ charges of the field combinations.
        Once again, although the operators $\mathcal{O}_{EEH}$ cannot induce NSI by themselves, they will come to play a related role, as they induce charged lepton flavour violating transitions.

Strictly speaking, not all of the above operators are independent when the full flavor structure is taken into account, as 
 \begin{align}
 (\mathcal{O}_{LLH}^{{\bf 313}})^{\beta \delta}_{\alpha \gamma}
=
 (\mathcal{O}_{LLH}^{{\bf 133}})^{\delta \beta}_{\gamma \alpha}
 \, .
\end{align}
However, the expressions below will look much simpler if both operators are used.

Notice that the phenomenologically interesting $\onsi$ operator in Eq.~(\ref{equ:nsi}) can be expressed as a combination of the two first operators in the list above,
\begin{equation}
\onsi= \frac{1}{2}\,(\mathcal{O}_{LEH}^{\bf 1}\,+\,\mathcal{O}_{LEH}^{\bf 3})\,.
\label{nsicomb}
\end{equation}
This means for instance that if a model only induces  at $d=8$  the operators $\mathcal{O}_{LEH}^{\bf 1}$ and $\mathcal{O}_{LEH}^{\bf 3}$ with similar weights and no $d=6$ NSI operator, it could be an optimal candidate for viable large NSI.
We will explore later some examples of this kind.
 
\end{itemize}

\subsubsection*{Decomposition in terms of $\boldsymbol{SU(2)}$ field components}

After EWSB, the contributions from the $d=6$ and $d=8$ gauge invariant operators result in two very simple sets of operators.
 From  the $\bar{L}L\bar{E}E$-type operators,
 Eqs.~(\ref{equ:ole}),~(\ref{equ:br1}) and~(\ref{equ:br2}),
 we find: 
 \begin{align}
\delta \mathscr{L}_{\text{eff}}
=&
 \frac{1}{\Lambda^{2}}
 \left(
 -
 \frac{1}{2} 
 \mathcal{C}_{LE}
 +
\frac{v^{2}}{2\Lambda^{2}}( \mathcal{C}_{LEH}^{{\bf 1}} 
 +
 \mathcal{C}_{LEH}^{\bf 3})
 \right)_{\beta \delta}^{\alpha \gamma}
 \left(
 \bar{\nu}^{\beta} \gamma^{\rho} {\rm P}_{L} \nu_{\alpha}
 \right)
 \left(
 \bar{\ell}^{\delta}
 \gamma_{\rho} {\rm P}_{R}
 \ell_{\gamma}
 \right) \nonumber \\
 & +
 \frac{1}{\Lambda^{2}}
 \left(
 -
 \frac{1}{2}
 \mathcal{C}_{LE}
 +
\frac{v^{2}}{2 \Lambda^{2}}( \mathcal{C}_{LEH}^{{\bf 1}} 
 -
 \mathcal{C}_{LEH}^{\bf 3})
 \right)_{\beta \delta}^{\alpha \gamma}
 \left(
 \bar{\ell}^{\beta} \gamma^{\rho} {\rm P}_{L} \ell_{\alpha}
 \right)
 \left(
 \bar{\ell}^{\delta}
 \gamma_{\rho} {\rm P}_{R}
 \ell_{\gamma}
 \right) +
 {\rm h.c.}\,,
\label{equ:LLEE}
\end{align}
 The first line in this equation produces the relevant NSI, whereas the second line leads to the (unwanted)
four charged lepton contributions. The NSI in the first line
involve only right-handed charged leptons. In consequence, their
effect at the neutrino source  will be chirally suppressed\footnote{ At detection, the effect of these NSI is subdominant because of the dominance of the neutrino-nucleon cross section.}.

 From the operators involving four lepton doublets,  Eqs.~(\ref{equ:oll1}),~(\ref{equ:oll3}), (\ref{equ:br3})-(\ref{equ:br7}),
 it results that~\footnote{Here we do not show the interactions among four neutrinos which these operators also induce.
See \App~\ref{app:fournu} for a discussion of these interactions.}
\begin{align}
\delta \mathscr{L}_{\text{eff}}
&= 
 \frac{1}{\Lambda^{2}}
 \left( 
 \mathcal{C}_{\mathrm{NSI}}^{\bar{L}L\bar{L}L} 
 \right)^{\alpha \gamma}_{\beta \delta} 
 \left(
 \bar{\nu}^{\beta} \gamma^{\rho} {\rm P}_{L} \nu_{\alpha}
  \right) \, 
 \left(
 \bar{\ell}^{\delta} \gamma^{\rho} {\rm P}_{L} \ell_{\gamma} 
 \right)
 \nonumber \\
 &+
 \frac{1}{\Lambda^{2}}
 \left(
 \mathcal{C}_{LL}^{{\bf 1}}
 +
 \mathcal{C}_{LL}^{{\bf 3}}
 +
 \frac{v^{2}}{2\Lambda^{2}} \left(
 \mathcal{C}_{LLH}^{{\bf 111}}
 +
 \mathcal{C}_{LLH}^{{\bf 331}}
 -
 \mathcal{C}_{LLH}^{{\bf 133}}
 -
 \mathcal{C}_{LLH}^{{\bf 313}} \right)
 \right)_{\beta \delta}^{\alpha \gamma}
 \left(
 \bar{\ell}^{\beta} \gamma^{\rho} {\rm P}_{L} \ell_{\alpha}
 \right)
 \left(
 \bar{\ell}^{\delta} \gamma^{\rho} {\rm P}_{L} \ell_{\gamma} 
 \right) \nonumber \\
&+ \mathrm{h.c.} \,, \label{equ:LLLL}
\end{align}
where
\begin{align}
\left(
\mathcal{C}_{\mathrm{NSI}}^{\bar{L}L\bar{L}L}
\right)^{\alpha \gamma}_{\beta \delta}
&= \left(
 \mathcal{C}_{LL}^{{\bf 1}}
 -
 \mathcal{C}_{LL}^{{\bf 3}}
 +
 \frac{v^{2}}{2\Lambda^{2}} \left(
 \mathcal{C}_{LLH}^{{\bf 111}}
 -
 \mathcal{C}_{LLH}^{{\bf 331}}
 -
 \mathcal{C}_{LLH}^{{\bf 133}}
 +
 \mathcal{C}_{LLH}^{{\bf 313}}\right)
 \right)^{\alpha \gamma}_{\beta \delta}
 \nonumber \\
 &\hspace{0.5cm}
+ \left(
 \mathcal{C}_{LL}^{{\bf 1}}
 -
 \mathcal{C}_{LL}^{{\bf 3}}
 +
 \frac{v^{2}}{2\Lambda^{2}} \left(
 \mathcal{C}_{LLH}^{{\bf 111}}
 -
 \mathcal{C}_{LLH}^{{\bf 331}}
 +
 \mathcal{C}_{LLH}^{{\bf 133}}
 -
 \mathcal{C}_{LLH}^{{\bf 313}}\right)
 \right)^{\gamma \alpha}_{\delta \beta}
\nonumber \\
 &\hspace{0.5cm}
 + \left(
 2 \, \mathcal{C}_{LL}^{{\bf 3}}
 + 
 \frac{v^{2}}{\Lambda^{2}}
 \left( \mathcal{C}_{LLH}^{{\bf 331}}
 - 
 \mathcal{C}_{LLH}^{{\bf 333}} \right)
 \right)_{\delta \beta}^{\alpha \gamma} 
\nonumber \\ 
 &\hspace{0.5cm}
 + \left(
 2 \, \mathcal{C}_{LL}^{{\bf 3}}
 +
 \frac{v^{2}}{\Lambda^{2}}
 (\mathcal{C}_{LLH}^{{\bf 331}}
 + 
 \mathcal{C}_{LLH}^{{\bf 333}})
\right)_{\beta \delta}^{\gamma \alpha} \, .
\label{equ:LLLLonsi}
\end{align}
Note the different flavor structure in the four lines in
Eq.~\eqref{equ:LLLLonsi}. In addition, note that the term relevant
for the NSI, \ie, the first line in \equ{LLLL}, couples to left-handed
charged leptons, which means that source NSI can be
generated as well. 
In resume, matter NSI are (not) correlated with source and production NSI for $\bar{L}L\bar{L}L$ ($\bar{L}L\bar{E}E$)-type operators.

\subsubsection*{Connection to NSI and phenomenology}

Let us first consider NSI in matter. The phenomenology  of neutrino propagation  under these conditions is customarily 
 described in terms of the Hamiltonian in the flavour basis~\cite{Grossman:1995wx, Campanelli:2002cc, Davidson:2003ha}, 
\begin{eqnarray}
 \mathcal{H}_F & = &
 \frac{1}{2E}
 \left\{
 U
 \begin{pmatrix}
  0 && \\
  & \Delta m_{21}^{2} & \\
  && \Delta m_{31}^{2}
 \end{pmatrix}
 U^{\dagger} + 
  a_{\rm CC}
 \begin{pmatrix}
  1+ \epsilon^{m}_{ee} &  \epsilon^{m}_{e\mu} &  \epsilon^{m}_{e\tau} \\
   (\epsilon^{m}_{e\mu})^{*}  &  \epsilon^{m}_{\mu\mu} 
  &  \epsilon^{m}_{\mu\tau} \\
   (\epsilon^{m}_{e\tau})^{*} &  (\epsilon^{m}_{\mu\tau})^{*} &  
  \epsilon^{m}_{\tau\tau}
 \end{pmatrix}
 \right\} \, ,
\label{equ:Ham}
\end{eqnarray}
where, $a_{\rm CC}$ is the usual matter effect term defined 
as $a_{\rm CC} \equiv 2 \sqrt{2} E G_{F} N_{e}$ (with $N_e$ the electron number density in
Earth matter). 

 From Eqs.~(\ref{equ:LLEE}) and (\ref{equ:LLLL}) it follows that
\be
\epsilon^{m,L}_{\beta \alpha}
 =
 \frac{v^{2}}{2\Lambda^{2}}
\left(
\mathcal{C}_{\mathrm{NSI}}^{\bar{L}L\bar{L}L}
\right)^{\alpha e}_{\beta e}
 \, ,  \quad 
\epsilon^{m,R}_{\beta \alpha}
 =
 \frac{v^{2}}{2\Lambda^{2}}
 \left(
 -
 \frac{1}{2}
 \mathcal{C}_{LE}
 +
 \frac{v^{2}}{2\Lambda^{2}}(\mathcal{C}_{LEH}^{{\bf 1}} 
 +
 \mathcal{C}_{LEH}^{\bf 3} )
 \right)_{\beta e}^{\alpha e}\,,
\label{equ:em}
\ee
with $\mathcal{C}_{\mathrm{NSI}}^{\bar{L}L\bar{L}L}$ as defined in \equ{LLLLonsi}. These two parameters in matter lead to  a total
\begin{equation}
\epsilon^{m}_{\beta \alpha} = \epsilon^{m,L}_{\beta \alpha} +  \epsilon^{m,R}_{\beta \alpha}\,, 
\end{equation}
because matter effects are only sensitive to the vector component. 

In addition to the propagation in matter, the production or detection processes can be affected by NSI. For the specific case of a neutrino factory and considering just the purely leptonic NSI under discussion, only effects at the source are relevant, since the detection interactions involve quarks~\footnote{Superbeams, for instance, use hadronic interactions for neutrino production, which are not affected by purely leptonic NSI to first order.}. They are customarily parametrized in terms of $\epsilon_{\alpha \beta}^s$, which describes an effective source state $ |\nu^s_\alpha \rangle$ as~\cite{Grossman:1995wx, Gonzalez-Garcia:2001mp, Bilenky:1992wv} 
\be \label{equ:nsisource}
|\nu^s_\alpha \rangle = |\nu_\alpha\rangle + \sum_{\gamma=e,\mu,\tau}\epsilon^s_{\alpha\gamma} |\nu_\gamma\rangle \, .
\ee 
 In this case, the muon decay rate could be modified by the NSI interaction in \equ{LLLL}, with the largest effect resulting from the coherent contribution to the state at the source~\cite{Gonzalez-Garcia:2001mp, Huber:2002bi}.
It appears as an admixture of a given flavour $\nu_\alpha$ with all other flavours, encoded by $\nu_\gamma$ in Eq.~(\ref{equ:nsisource}).  
 Two types of contributions are possible, 
\be
\epsilon^{s}_{\mu \beta} 
 =
 \frac{v^{2}}{2\Lambda^{2}}
(\mathcal{C}_{\mathrm{NSI}}^{\bar{L}L\bar{L}L})^{e \mu}_{\beta e} \quad \text{or} \quad
\epsilon^{s}_{e \beta} 
 =
 \frac{v^{2}}{2\Lambda^{2}}
(\mathcal{C}_{\mathrm{NSI}}^{\bar{L}L\bar{L}L})^{\mu e}_{\beta \mu}\,.
\label{equ:nsisource2}
\ee
The second possibility will affect the golden $\nu_e \rightarrow \nu_\mu$ appearance channel, where the effect might be easiest to observe.
If the coefficients in \equ{LLLLonsi} are known for a specific model, one can easily calculate  the connection between source and propagation effects via Eqs.~(\ref{equ:em}) and (\ref{equ:nsisource2}), a connection which does not hold for $\bar{L}L \bar{E} E$-type operators above,  as explained earlier.

\subsubsection*{Conditions to suppress charged lepton processes }

Let us discuss now potentially dangerous contributions to charged lepton
  flavour violation processes,  possible modifications of $G_F$ and the
  constraints on lepton universality.  The focus is set on pure
  charged lepton processes at tree level.
  These interactions can result from the
  second terms in Eqs.~(\ref{equ:LLEE}) and (\ref{equ:LLLL}). 
  They should be very suppressed in any phenomenologically viable model.  
  In order to cancel those terms, the putative beyond the SM theory has to satisfy, to a high degree of accuracy, the following constraints:
\begin{align}
\label{equ:cancel1}
\left(
 -
 \frac{1}{2}
 \mathcal{C}_{LE}
 +
\frac{v^{2}}{2 \Lambda^{2}}( \mathcal{C}_{LEH}^{{\bf 1}} 
 -
 \mathcal{C}_{LEH}^{\bf 3})
 \right)_{\beta \delta}^{\alpha \gamma} & = 0 \, , \\
 \left(
 \mathcal{C}_{LL}^{{\bf 1}}
 +
 \mathcal{C}_{LL}^{{\bf 3}}
 +
 \frac{v^{2}}{2\Lambda^{2}} \left(
 \mathcal{C}_{LLH}^{{\bf 111}}
 +
 \mathcal{C}_{LLH}^{{\bf 331}}
 -
 \mathcal{C}_{LLH}^{{\bf 133}}
 -
 \mathcal{C}_{LLH}^{{\bf 313}} \right)
 \right)_{\beta \delta}^{\alpha \gamma} & = 0\,,
\label{equ:cancel2}
\end{align}
for all possible values of the flavour indices (Greek letters).
A possibility suggested by these equations is that 
 there could be cancellations among $d=6$  and $d=8$ operator coefficients.
 However, we will not discuss such a possibility in this study, as it would 
correspond to fine-tune the scale $\Lambda$. 
We will therefore require that
the $d=6$ and $d=8$ operator coefficients in \equ{cancel1} and \equ{cancel2}
cancel independently. 

For the $d=6$ operator coefficients, it reads (omitting flavor indices)
\be
 \mathcal{C}_{LE} = 0 \, , \quad  \mathcal{C}_{LL}^{{\bf 1}}
 = - \mathcal{C}_{LL}^{{\bf 3}}, \, , \quad \mathcal{C}_{EE} = 0 \,,
\label{equ:cancelsix}
\ee
which implies that only  $\bar{L}L\bar{L}L$-type operators can induce large NSI.
One possibility for its implementation is the antisymmetric operator mediated by a $SU(2)$ singlet scalar in \Ref~\cite{Bergmann:1999pk},
which turns out to be the only $d=6$ possibility requiring just one tree-level mediator, as we shall explicitely demonstrate.

For the $d=8$ operator coefficients, the cancellation conditions read 
\be
 \mathcal{C}_{LEH}^{{\bf 1}} = \mathcal{C}_{LEH}^{\bf 3} \, , \quad \mathcal{C}_{LLH}^{{\bf 111}}
 +
 \mathcal{C}_{LLH}^{{\bf 331}}
 -
 \mathcal{C}_{LLH}^{{\bf 133}}
 -
 \mathcal{C}_{LLH}^{{\bf 313}} = 0 \, , \quad
 \mathcal{C}_{LLH}^{{\bf 333}} \, \, \text{arbitr.} \, , \quad
 \mathcal{C}_{EEH} = 0 \,,
\label{equ:canceleight}
\ee
where the first condition corresponds to operators of the type $\onsi$ in \equ{nsi}, see Eq.~(\ref{nsicomb}).
In the following, we will refer to operators satisfying \equ{canceleight} as $\onsi$, \ie, we define the
class of potential non-standard neutrino interaction operators in  mass dimension eight as the one which does not introduce any  harmful $d=8$ processes with four charged leptons. \equ{nsi} is (apart from Fierz rearrangements) the only such possibility with two right-handed charged leptons involved. When considering leptonic NSI involving four left-handed fields, several new operators of this kind will be determined later on.

As far as the possible NSI in terms of $SU(2)$ field components are
concerned, not all  flavour structures can be generated from the $d=6$  effective gauge invariant operators  if charged lepton processes are suppressed. Applying the $d=6$ cancellation conditions in \equ{cancelsix} to 
\equ{LLLLonsi},  it results that the $d=6$ contribution to the coefficient $\left(\mathcal{C}_{\mathrm{NSI}}^{\bar{L}L\bar{L}L} 
 \right)^{\alpha \gamma}_{\beta \delta}$ is  antisymmetric in the flavor index exchanges $(\alpha,\gamma) \rightarrow (\gamma,\alpha)$ and
$(\beta, \delta) \rightarrow (\delta,\beta)$, which means that $\alpha \neq \gamma$ and $\beta \neq \delta$ for
viable NSI. As regards matter effects, this implies that only $\epsilon^m_{\mu\mu}$, $\epsilon^m_{\mu\tau}$, and $\epsilon^m_{\tau\tau}$ -- defined in 
Eq.~(\ref{equ:em}) -- can be generated from $d=6$ operators and the connection with the source effects  is given by
\begin{equation}
\epsilon^m_{\mu \mu} = - \epsilon^s_{ee}= - \epsilon^s_{\mu\mu}\,,\label{ay}
\end{equation}
\begin{equation}
 \epsilon^m_{\mu \tau} = - (\epsilon^s_{\mu \tau})^{*}\,.
\label{ay2}
\end{equation}
In contrast, $\epsilon^m_{\tau \tau}$ is not connected to the source
effects at the effective operator level~\footnote{There can be also
subdominant effects in detection chains. 
For example, in OPERA, the taus resulting from hadronic interactions
decay into muons or electrons. 
It implies for instance $\epsilon^m_{\tau \tau}= - \epsilon_{\tau
\tau}^s$, which means that tau decay into electrons is in this case
connected with matter NSI. Note that NSI and SM contributions add coherently to the $\tau^{-}
\rightarrow e^{-} \nu_{\tau} \bar{\nu}_{e}$ width.}. 
 Notice that, for instance, the NSI in Eq.~(\ref{ay})  contribute to the
$G_F$ measurement coherently (\ie, the interference with SM couplings contributes linearly to the rates), for which quite stringent bounds exist. 
These results hold in general for any  purely leptonic NSI $d=6$ operator with suppressed interactions among four charged leptons, \ie, \equ{cancelsix}.  Furthermore, 
for the particular case of a neutrino factory, the antisymmetry conditions  and constraints described above imply that the only possible non-negligible NSI source terms induced by $d=6$ operators  are 
$\epsilon^s_{e \tau}$ and $\epsilon^s_{\mu \tau}$.

\section{Model analysis of $\boldsymbol{d=6}$ operators}
\label{sec:six}

\begin{table}[tp]
\begin{center}
\small{
\begin{tabular}{ccclccl}
\hline \hline
$\Delta L$ 
 & $SU(2)_{L}$
 & $U(1)_{Y}$
 & Lorentz
 & Mediator
 & Bilinear(s)
 & Models [Refs.] \\
\hline
 2
 & 
 {\bf 1}
 & 
 $-1$
 & 
 scalar
 &
 ${\bf 1}_{-1}^s$
 &
 $
 \overline{L^{c}} {\rm i}\tau^{2} L$
 &
 Zee model~\cite{Zee:1980ai,Zee:1985rj,Babu:2002uu}, $\Slash{R_{p}}$SUSY~\cite{Barbier:2004ez} \\
 &
 & $-2$
 & scalar
 &
 ${\bf 1}_{-2}^s$
 &
 $\overline{E^{c}} E$
 &
 \\
 & {\bf 3}
 & $-1$
 & scalar
 &
 ${\bf 3}_{-1}^s$
 &
 $\overline{L^{c}} {\rm i} \tau^{2} \tau^{a} L$
 &
 Left-right sym.~\cite{Pati:1974yy,Mohapatra:1974gc,Senjanovic:1975rk,Mohapatra:1980yp}
 \\
 & {\bf 2}
 & $-3/2$
 & vector
 &
 ${\bf 2}_{-3/2}^v$
 &
 $\overline{E^{c}} \gamma^{\rho} L$
 &
 331 model~\cite{Pisano:1991ee,Frampton:1992wt,Foot:1992rh}
 \\
\hline
 0
 & {\bf 1}
 & 0
 & vector 
 &
 ${\bf 1}_{0}^v$
 &
 $\bar{L} \gamma^{\rho} L$,  $\bar{E} \gamma^{\rho} E$
 &
 Models with $Z'$~\cite{Yao:2006px}\\
 & {\bf 3}
 & 0
 & vector 
 &
 ${\bf 3}_{0}^v$
 &
 $\bar{L} \gamma^{\rho}  \tau^{a} L$
 &
 Models with $W'$~\cite{Yao:2006px}
 \\
 & {\bf 2}
 & 1/2
 & scalar &
 ${\bf 2}_{1/2}^s$
 &
 $\bar{E}  L $
 &
 $\Slash{R_{p}}$SUSY~\cite{Barbier:2004ez}\\
 \hline \hline
\end{tabular}
} 
\end{center}
\mycaption{\label{tab:bil} Possible SM bilinear field combinations involving only leptons.  $\Delta L$ refers to the lepton number of the mediator, 
$SU(2)_L$ to electroweak singlets ($\boldsymbol{1}$), doublets ($\boldsymbol{2}$), or triplets ($\boldsymbol{3}$), $U(1)_Y$ to the hypercharge $Y \equiv Q- I^W_3$, and ``Lorentz'' to the Lorentz nature. 
The mediator notation is $\boldsymbol{X}^{\mathcal{L}}_Y$, where $\boldsymbol{X}$, $\mathcal{L}$ and  $Y$ denote its $SU(2)$, Lorentz, and $Y$  properties, respectively. The mediators which carry two units of lepton number were studied in Ref.~\cite{Cuypers:1996ia}.} 
\end{table}

\begin{table}[tbp]
\begin{center}
\begin{tabular}{lccccc}
\hline\hline
 $d=6$ operators 
&
Mediator
&
$\mathcal{C}_{LE} $
&
$\mathcal{C}_{LL}^{{\bf 1}}$
&
$\mathcal{C}_{LL}^{{\bf 3}}$
&
$\mathcal{C}_{EE}$
\\
\hline 
\multicolumn{6}{l}{$\boldsymbol{\bar{L}E\bar{E}L}$} \\
$
(c^{{\bf 2}v}/\Lambda^{2})
((\overline{ E^c})_\gamma \gamma^{\rho} L_\alpha)(\bar{L}^\beta \gamma_{\rho} (E^c)^\delta)$
&
${\bf 2}_{-3/2}^{v}$
&
$2c^{{\bf 2}v}$
&
&
&
\\
$(f_{LE}^{{\bf 1}v}/\Lambda^{2})(\bar{L}^\beta \gamma^{\rho} L_\alpha)(\bar{E}^\delta \gamma_{\rho} E_\gamma)$
&
${\bf 1}_{0}^{v}$
&
$-2f_{LE}^{{\bf 1}v}$
&
&
&
\\
$(f^{{\bf 2}s}/\Lambda^{2}) (\bar{L}^\beta E_\gamma)(\bar{E}^\delta L_\alpha)$
&
${\bf 2}_{1/2}^{s}$
&
$f^{{\bf 2}s}$
&
&
\\
\hline
\multicolumn{6}{l}{$\boldsymbol{\bar{L}L\bar{L}L}$} \\
$(c^{{\bf 1}s}_{LL}/\Lambda^{2})
((\overline{L^c})_\alpha i \tau^2 L_\gamma)(\bar{L}^\beta i \tau^2 (L^c)^\delta)$
&
${\bf 1}_{-1}^{s}$
&
&
$\frac{1}{4} c^{{\bf 1}s}$
&
$-\frac{1}{4} c^{{\bf 1}s}$
&
\\
$
(c^{{\bf 3}s}/\Lambda^{2})
((\overline{L^c})_\alpha i \tau^2 \vec\tau L_\gamma)(\bar{L}^\beta \vec\tau i \tau^2 (L^c)^\delta)$
&
${\bf 3}_{-1}^{s}$
&
&
$-\frac{3}{4} c^{{\bf 3}s}$
&
$-\frac{1}{4} c^{{\bf 3}s}$
&
\\
$(f^{{\bf 1}v}_{LL}/\Lambda^{2})
(\bar{L}^\beta \gamma^{\rho} L_\alpha) (\bar{L}^\delta \gamma_{\rho} L_\gamma)$
&
${\bf 1}_{0}^{v}$
&
&
$f^{{\bf 1}v}_{LL}$
&
&
\\
$(f^{{\bf 3}v}/\Lambda^{2})(\bar{L}^\beta \gamma^{\rho} \vec\tau L_\alpha) 
(\bar{L}^\delta \gamma_{\rho}\vec\tau L_\gamma)$
&
${\bf 3}_{0}^{v}$
&
&
&
$f^{{\bf 3}v}$
&
\\
\hline 
\multicolumn{6}{l}{$\boldsymbol{\bar{E}E\bar{E}E}$} \\
$(c^{{\bf 1}s}_{EE}/\Lambda^{2})((\overline{E^c})_\alpha  E_\gamma )(\bar{E}^\beta (E^c)^\delta)$ 
&
${\bf 1}_{-2}^{s}$
&
&
&
&
$\frac{1}{2}c^{{\bf 1}s}_{EE}$
\\
$(f^{{\bf 1}v}_{EE}/\Lambda^{2})(\bar{E}^\beta \gamma^{\rho}E_\alpha )(\bar{E}^\delta \gamma_{\rho} E_\gamma)$ 
&
${\bf 1}_{0}^{v}$
&
&
&
&
$f^{{\bf 1}v}_{EE}$
\\
\hline\hline
\end{tabular}
\end{center}
\mycaption{
\label{tab:opdim6}
 Possible $d=6$ operators obtained by combining the lepton bilinears in \Tab~\ref{tab:bil}.  The coefficients are labeled $c^{{\bf X}\mathcal{L}}$ ($\Delta L=2$) or $f^{{\bf X}\mathcal{L}}$ ($\Delta L=0$), where the ${\bf X}$ and $\mathcal{L}$  are defined as in Table 1 and the subscripts refer to the combination of bilinears, in an obvious notation.
 The last four columns contain their contribution to the $d=6$ operator coefficients in the BW basis in Eqs.~(\ref{equ:ole})-(\ref{equ:oee}).  The flavour structure for any coefficient in the table is understood to be $(\,)^{\alpha\gamma}_{\beta\delta}$, see main text for further explanations.} 
\end{table}

In this section, we discuss the model-building implications of requesting large $d=6$ NSI induced by theories of
physics beyond the Standard Model.  We specifically highlight the basic principles, which can be found in the $d=8$ case as well. However, as we shall see later, the $d=8$ case is technically somewhat more
challenging.

In order to shed light on model building, let us analyze the operators according to the possible tree-level mediator. 
This is most efficiently done by listing all possible SM bilinear field combinations, and combining them in all possible ways~\cite{Bergmann:1998ft,Bergmann:1999pk}.

We therefore show in \Tab~\ref{tab:bil} the possible bilinears constructed from leptons only, which can lead to the $d=6$ NSI operators in Eqs.~(\ref{equ:ole})-(\ref{equ:oee}). It is obvious from the table that the bilinears carry the mediator information and that they can therefore  be directly associated with specific models (as illustrated). 
The mediators are denoted -- all through the paper --  by  ${\bf X}^{\mathcal{L}}_Y$, where
\begin{itemize}
\item
 ${\bf X}$ denotes the $SU(2)$ nature, \ie, singlet ${\bf 1}$, doublet ${\bf 2}$, or triplet ${\bf 3}$.
 \item
 $\mathcal{L}$ refers to the Lorentz nature, \ie, scalar ($s$), vector ($v$),  left-handed ($L$) or right-handed ($R$) fermion~\footnote{ Fermionic mediators will appear explicitly later on, when discussion $d=8$ effective interactions}.
\item
 $Y$ refers to the hypercharge $Y=Q-I^W_3$.
\end{itemize}
 \Tab~\ref{tab:opdim6} shows in turn all possible $d=6$ operators which can be constructed from the SM bilinear field combinations in \Tab~\ref{tab:bil}. 
The coefficients of the $d=6$ operators obtained by this procedure are denoted  by 
 $(c^{{\bf X}\,\mathcal{L}})^{\alpha\gamma}_{\beta\delta}$ and $(f^{{\bf X}\,\mathcal{L}})^{\alpha\gamma}_{\beta\delta}$, where $c$ ($f$) indicates that the corresponding operator results from the exchange of particles carrying two (zero) lepton number, and ${\bf X}$ and $\mathcal{L}$ refer again to the $SU(2)$ and Lorentz nature, respectively. Any subscript refers to the combination of bilinears involved.

At this point it is important to note that the operators obtained from the mediators do not constitute a basis. Instead they are not independent, but  linear combinations of those in the BW basis, Eqs.~(\ref{equ:ole})-(\ref{equ:oee}). Therefore, it might be more accurate to call them ``mediator-operators'' or ``operator combinations''. We will not make this special distinction, but the reader should keep that in mind. 
Re-writing the individual effective operators from   \Tab~\ref{tab:opdim6} in the BW basis, we find the coefficients given in the last four columns of \Tab~\ref{tab:opdim6}. For example,  the first line of the second group, mediated by ${\bf 1}_{-1}^s$, reads (including flavor indices)
\be
\delta \mathscr{L}_{\text{eff}}^{d=6} =
\frac{(c^{{\bf 1}s}_{LL})^{\alpha \gamma}_{\beta \delta}}{\Lambda^{2}} \,
((\overline{L^c})_\alpha i \tau^2 L_\gamma)(\bar{L}^\beta i \tau^2
(L^c)^\delta) 
=  \frac{1}{4} \frac{(c^{{\bf 1}s}_{LL})^{\alpha \gamma}_{\beta
\delta}}{\Lambda^{2}} \, (\mathcal{O}^{\bf{1}}_{LL})^{\beta
\delta}_{\alpha \gamma} 
-\frac{1}{4} \frac{(c^{{\bf 1}s}_{LL})^{\alpha \gamma}_{\beta \delta}}{\Lambda^{2}} \, (\mathcal{O}^{\bf{3}}_{LL})^{\beta \delta}_{\alpha \gamma} \, . 
\label{equ:example1}
\ee
Conversely, the decomposition of the operator $\mathcal{O}^{\bf{1}}_{LL}$ of the BW basis can be read off from the column labeled $\mathcal{C}^{\bf{1}}_{LL}$, in terms of the relative weights of the mediator-operators:
\be
(\mathcal{C}_{LL}^{{\bf 1}})^{\alpha \gamma}_{\beta \delta}
=
\frac{1}{4} 
(c^{{\bf 1}s}_{LL})^{\alpha \gamma}_{\beta \delta}
-
\frac{3}{4} 
(c^{{\bf 3}s})^{\alpha \gamma}_{\beta \delta}
+
(f^{{\bf 1}v}_{LL})^{\alpha \gamma}_{\beta \delta}\,. 
\label{equ:example2}
\ee
Note that the flavor indices in the first column of \Tab~\ref{tab:opdim6} are arranged such that the flavor indices of all coefficients and of the BW operators are the same as in Eqs.~(\ref{equ:example1}) and~(\ref{equ:example2}). Therefore, we show the flavor indices explicitly only in the first column.

In order to have large NSI without four charged lepton interactions,
the $d=6$ cancellation conditions \equ{cancelsix} must now be implemented. 
One can directly read off now from \Tab~\ref{tab:opdim6} that these conditions can be re-written as
\begin{align}
\label{equ:cond1}
 2 c^{{\bf 2}v}
 -
 2 f^{{\bf 1}v}_{LE}
 +
f^{{\bf 2}s}
 &=
 0 \qquad (\text{from} \, \, \mathcal{C}_{LE}=0 )\, ,
 \\
 -
 c^{{\bf 3}s}
 +
 f^{{\bf 1}v}_{LL}
 +
 f^{{\bf 3}v}
 &=
 0 \qquad (\text{from} \, \, \mathcal{C}_{LL}^{{\bf 1}}  + \mathcal{C}_{LL}^{{\bf 3}} = 0) \, ,
\label{equ:cond2} \\
c^{{\bf 1}s}_{EE} + 2 f^{{\bf 1}v}_{EE} & = 0
\qquad (\text{from} \, \, \mathcal{C}_{EE} = 0)\,,
\label{equ:cond3}
\end{align}
in the mediator picture. The operators contributing to the first
equation will not produce any NSI (since $\mathcal{C}_{LE}=0$ in
\equ{LLEE}), while the operators present in the second equation lead to
NSI if  $\mathcal{C}_{LL}^{{\bf 1}} = - \mathcal{C}_{LL}^{{\bf 3}} \neq 0$ (\cf,  \equ{LLLL}).

One approach to use \Tab~\ref{tab:opdim6} is to discuss departures from
the SM couplings. For example, for a hypothetical experimental departure pointing towards
 a four-lepton coupling such as that in operator $\mathcal{O}_{LL}^{{\bf 3}}$ in
\equ{oll3},  \Tab~\ref{tab:opdim6} indicates directly that a new heavy scalar triplet could induce it at
tree-level, while a scalar doublet wouldn't.

 From the model building
perspective,  it is illustrative to consider again the case of  the operator mediated by 
${\bf 1}_{-1}^s$  
leading to \equ{example1}. The table shows that it is the only $d=6$ possibility using only one mediator which directly
satisfies the cancellation condition of pure charged lepton interactions \equ{cancelsix} (or their tree-level equivalent Eqs.~(\ref{equ:cond1})-(\ref{equ:cond3}) ). 
This antisymmetric combination of the basis elements was first found in
\Ref~\cite{Bergmann:1999pk}. 

\begin{figure}[t]
\unitlength=0.9cm
\begin{picture}(5,3)
\thicklines
 \put(0.5,0.5){\line(1,0){4}}
 \put(0.5,2.5){\line(1,0){4}}
 \multiput(2.5,2.5)(0,-0.3){7}{\line(0,-1){0.2}}
 \put(2.5,0.5){\circle*{0.2}}
 \put(2.5,2.5){\circle*{0.2}}
 \put(1.5,2.5){\vector(1,0){0}}
 \put(3.5,2.5){\vector(-1,0){0}}
 \put(1.3,0.5){\vector(-1,0){0}}
 \put(3.7,0.5){\vector(1,0){0}}
 \put(2.5,2.7){$\lambda_{e \mu}$}
 \put(2.5,0.05){$(\lambda_{e \tau})^{*}$}
 \put(2.7,1.3){$1/M_{{\bf 1}s}^{2}$}
 \put(0.3,0.2){$L_{e}$}
 \put(4.5,0.2){$L_{\tau}$}
 \put(0.3,2.7){$L_{e}$}
 \put(4.5,2.7){$L_{\mu}$}
\end{picture}
\hspace{1cm}
\begin{picture}(5,3)
\thicklines
 \put(0.5,0.5){\line(1,0){4}}
 \put(0.5,2.5){\line(1,0){4}}
 \multiput(2.5,2.5)(0,-0.3){7}{\line(0,-1){0.2}}
 \put(2.5,0.5){\circle*{0.2}}
 \put(2.5,2.5){\circle*{0.2}}
 \put(1.5,2.5){\vector(1,0){0}}
 \put(3.5,2.5){\vector(-1,0){0}}
 \put(1.3,0.5){\vector(-1,0){0}}
 \put(3.7,0.5){\vector(1,0){0}}
 \put(2.5,2.7){$\lambda_{e \mu}$}
 \put(2.5,0.05){$(\lambda_{e \mu})^{*}$}
 \put(2.7,1.3){$1/M_{{\bf 1}s}^{2}$}
 \put(0.3,0.2){$L_{e}$}
 \put(4.5,0.2){$L_{\mu}$}
 \put(0.3,2.7){$L_{e}$}
 \put(4.5,2.7){$L_{\mu}$}
\end{picture}
\hspace{1cm}
\begin{picture}(5,3)
\thicklines
 \put(0.5,0.5){\line(1,0){4}}
 \put(0.5,2.5){\line(1,0){4}}
 \multiput(2.5,2.5)(0,-0.3){7}{\line(0,-1){0.2}}
 \put(2.5,0.5){\circle*{0.2}}
 \put(2.5,2.5){\circle*{0.2}}
 \put(1.5,2.5){\vector(1,0){0}}
 \put(3.5,2.5){\vector(-1,0){0}}
 \put(1.3,0.5){\vector(-1,0){0}}
 \put(3.7,0.5){\vector(1,0){0}}
 \put(2.5,2.7){$\lambda_{e \tau}$}
 \put(2.5,0.05){$(\lambda_{e \tau})^{*}$}
 \put(2.7,1.3){$1/M_{{\bf 1}s}^{2}$}
 \put(0.3,0.2){$L_{e}$}
 \put(4.5,0.2){$L_{\tau}$}
 \put(0.3,2.7){$L_{e}$}
 \put(4.5,2.7){$L_{\tau}$}
\end{picture}
\mycaption{Diagrams mediated by a bilepton ${\bf 1}^s_{-1}$.
 The effective dimension six operator results from the 
the first diagram at energies below the mediator mass $M_{\bf{1}s}$. However, the fundamental interaction will also create the diagrams in the middle and right, and the couplings will be related.}
\label{fig:bilepton-med}
\end{figure}

This example serves to illuminate the power of the mediator analysis (see \Ref~\cite{Bergmann:1999pk} and also \Ref~\cite{Antusch:2008tz}). The ${\bf 1}^s_{-1}$ exchange leading to the originally proposed operator is 
  depicted in \figu{bilepton-med}, left. Once a certain mediator is assumed for a certain operator, contributions to other operators are simultaneously induced, though, as illustrated at the center and right of \figu{bilepton-med}, 
  \ie, 
 \begin{align}
 \frac{\left|
 (c^{{\bf 1}s}_{ LL})^{e\mu}_{e\tau}
 \right|^{2}}{\Lambda^{4}}
 =
 \frac{|\lambda_{e\mu}|^{2} |\lambda_{e\tau}|^{2}}{M_{{\bf 1}s}^{4}} 
 =
\frac{
 \left|
 (c^{{\bf 1}s}_{ LL})^{e\mu}_{e\mu}
 \right|
 \left|
 (c^{{\bf 1}s}_{ LL})^{e\tau}_{e\tau}
 \right|}{\Lambda^{4}} \, ,
\label{equ:fund}
\end{align}
where $\lambda_{\alpha \beta}$ is the coupling for 
the lepton-bilepton interaction and $M_{{\bf 1} s}$ is the mass of 
the bilepton.  A  coherent contribution to $G_F$ and a violation of the lepton universality is 
then induced by the diagrams at the middle and right of \figu{bilepton-med}. From the strict experimental bounds on these quantities $\epsilon^m_{\mu \tau}$ has been constrained to $|\epsilon_{\mu \tau}^m| \lesssim 1.9 \cdot 10^{-3}$ (90\% CL), using this particular mediator~\cite{Antusch:2008tz}. The bound from a neutrino factory on $\emta$ would be
$1.8 \cdot 10^{-2}$ for complex $\epsilon^m_{\mu \tau}$~\cite{Kopp:2008ds}. If it was assumed to be real, which does not describe the most general class of models, the bound would be $3.7 \cdot 10^{-4}$ (90\% CL)~\cite{Kopp:2008ds}. However, since this is a model-dependent assumption, we do not use this bound.

 The antisymmetric operator discussed in the previous paragraphs  is not, however,  the only possibility to
build a model satisfying Eqs.~(\ref{equ:cond1})-(\ref{equ:cond3}).  
 For example, one may choose
bosonic triplets ${\bf 3}^s_{-1}$ and
${\bf 3}^v_0$, for which the coefficients can be chosen independently,
in order to satisfy \equ{cond2} without suppressing completely the $d=6$
NSI operator coefficient.  
In particular, if the simplest possibility is experimentally constrained, one may consider models with more than one mediator. 

At this point, we would like to clarify that cancellations or fine-tuning of operator coefficients cannot be an argument in itself for judging the naturalness and complexity of a model. It depends on the field content and the symmetries of the model.
Consider for instance once again the antisymmetric operator in the
left-hand side of Eq.~(\ref{equ:example1}), induced at tree-level by the
exchange of just one mediator, ${\bf 1}_{-1}^s$, illustrated in
Fig.~\ref{fig:bilepton-med}  left.  That equation shows that, in the BW basis, the antisymmetric operator  appears to be constructed from the combination of two BW operators with specific (fine-tuned?) coefficients.  In the effective operator picture, ``fine-tuning" is thus a basis and model-dependent qualification. 
We therefore define the simplest model to be the one with the fewest mediators. In the $d=6$ case, it is the antisymmetric operator in \equ{example1} with only one mediator. In the case that the NSI come only through $d=8$ (or higher dimension) effective operators, 
 we will demonstrate that the simplest viable models require at least two mediators.
Once the field content is chosen, any relative precise adjustment of the couplings of the mediators can be considered a fine-tuning, unless the symmetries of the model ensure it.  It will be left to the model builder to eventually explore possible symmetries, whenever such cancellations will turn out to be required below for phenomenologically viable NSI.

In general,  it is easy to show  that all NSI from $d=6$ operators are 
strongly constrained when the possible mediators are taken into account. 
There is, however, one exception. The present experimental constraints allow 
 the condition $\mathcal{C}_{LE}=0$ in \equ{cancelsix}, which
 cancels interactions among four charged leptons,
 to be substantially violated for certain combinations of flavour indices. 
In particular, the coefficient of the flavor conserving (BW) operator
$(\bar{L}^{\tau}E_{e})(\bar{E}^{e} L_{\tau} )$,
is not very strongly constrained~\cite{Berezhiani:2001rs,Ibarra:2004pe}.
 The mediators ${\bf 2}^{v}_{-3/2}$ or
${\bf 2}_{1/2}^{s}$ (\cf, \Tab~\ref{tab:opdim6}) can generate such an operator, leading to 
the following effective interactions, \cf, \equ{LLEE}:
\begin{align}
\delta\mathscr{L}_{\text{eff}}^{d=6}
=
-\frac{(\mathcal{C}_{LE})_{\tau e}^{\tau e}}{2 \Lambda^{2}}
\left(
 \left(
 \bar{\nu}^{\tau} \gamma^{\rho} {\rm P}_{L} \nu_{\tau}
 \right)
 \left(
 \bar{e} \gamma_{\rho} {\rm P}_{R} e
 \right)
-
 \left(
 \bar{\tau} \gamma^{\rho} {\rm P}_{L} \tau
 \right)
 \left(
 \bar{e} \gamma_{\rho} {\rm P}_{R} e
 \right)
 \right)
 +{\rm h.c.} \, .
\end{align}
The coefficient is constrained by (see Eq.(14) 
in Ref.\cite{Berezhiani:2001rs}) 
\begin{align}
|\epsilon^{m}_{\tau\tau}|
=
\frac{v^2}{4\Lambda^2} \, |(\mathcal{C}_{LE})_{\tau e}^{\tau e}|
=
|\kappa_{\tau R}| 
\lesssim 
0.1 \, .
\end{align}
If the possibility of large $SU(2)_{L}$ breaking effects was considered in addition,
a possible gain of almost an order of magnitude could be allowed for the NSI  $\epsilon^{m}_{\tau\tau}$ strength~\cite{Bergmann:1999pk}.
In conclusion,  large (order unity) values for $\epsilon^{m}_{\tau\tau}$ resulting from  $d=6$
effective interactions are not excluded.

 \Tab~\ref{tab:opdim6} also shows that the relationship between mediator and
coefficient is unique at the $d=6$ level, except for ${\bf 1}_{0}^{v}$. If a model uses this mediator,
then there will be three different $d=6$ operator contributions, which are independent in the BW
basis. In particular, one cannot neglect $\mathcal{O}_{EE}$,  which can
induce physics effects while not resulting in  NSI.

\section{Model analysis of $\boldsymbol{d=8}$  operators}
\label{sec:eight}

We consider all possible $d=8$ operators which can induce purely leptonic NSI, analyzing them from the point of view of their possible tree-level mediators. We will focus on the systematic analysis of all possible products of SM bilinears, which may result from exchanging mediators which only couple to pairs of SM fields. Such an 
 analysis was performed for $d=6$ operators in \Refs~\cite{Bergmann:1998ft,Bergmann:1999pk}, and we extend it here to the $d=8$ case. 
Other scenarios leading to some of the $d=8$ operators will be briefly analyzed afterwards.

A convenient basis of linearly independent $d=8$ operators has been given in
Eqs.~(\ref{equ:br1}) to~(\ref{equ:br8}), \ie, the BR basis. In order to
suppress four charged lepton interactions, both the cancellation conditions for  $d=8$ operators in \equ{canceleight} and the cancellation conditions for $d=6$ operators  in  \equ{cancelsix} are now required to be satisfied. 
 Under these conditions, if any $d=6$ operator remains, it is expected to dominate the new physics and, as discussed in the previous section,
  only effects related to $\epsilon^m_{\tau\tau}$ are then allowed to be experimentally sizeable. In this section, we instead focus on NSI which stem exclusively from $d=8$ (and higher) operators and their implications for model building. In particular, we are interested in the $\onsi$ without four charged lepton interactions, \ie, satisfying Eq.~(\ref{equ:canceleight}),  which has been object of intense speculations in the literature.  
  
  When the mediators couple only to SM bilinears we have the following options with respect to the  undesired $d=6$ operators: 
\begin{enumerate}
\item The required mediators do not induce any $d=6$ operator involving four leptons (in other words, the mediators differ from those in \Tab~\ref{tab:opdim6}).
\item The $d=6$ couplings induced by different mediators turn out to explicitly cancel among themselves.
\end{enumerate}
 As we will illustrate later, there is no simple possibility for which the first option works. For the second option to happen,
 the coefficients for the BW operators in Eqs.~(\ref{equ:ole})-(\ref{equ:oee}) have to vanish independently, because they constitute a basis:
\be
 \mathcal{C}_{LE}=0 \, , \quad  \mathcal{C}_{LL}^{{\bf 1}}  = 0 \, ,  \quad  \mathcal{C}_{LL}^{{\bf 3}} = 0 \, , \quad
\mathcal{C}_{EE}=0 \, . 
\label{equ:cancelall}
\ee
Their implementation in the mediator picture can be read off from the columns in \Tab~\ref{tab:opdim6}. They are given by Eqs.~(\ref{equ:cond1}) and 
(\ref{equ:cond3}), together with 
\begin{align}
\label{equ:cond2a}
 \frac{1}{4}c^{{\bf 1}s}_{ LL}
 -
 \frac{3}{4} c^{{\bf 3}s}
 +
f^{{\bf 1}v}_{LL}
 &=
 0 \qquad (\text{from} \, \, \mathcal{C}_{LL}^{{\bf 1}}=0 )\, ,
 \\
 -
 \frac{1}{4}c^{{\bf 1}s}_{ LL}
 -
 \frac{1}{4} c^{{\bf 3}s}
 +
 f^{{\bf 3}v}
 &=
 0 \qquad (\text{from} \, \, \mathcal{C}_{LL}^{{\bf 3}} = 0) \, ,
\label{equ:cond2b} 
\end{align}
which replace Eq.~(\ref{equ:cond2}) of that set.
 For example, if a model introduces 
two bosonic doublets ${\bf 2}_{-3/2}^v$ and ${\bf 2}_{1/2}^s$, one can satisfy \equ{cond1} (to which Eq.~(\ref{equ:cancelall}) simplifies in this case) 
by achieving $2 c^{{\bf 2}v}+f^{{\bf 2}s}=0$.

Note that the introduction of exotic fermions in the game potentially leads to the
additional $d=6$ operators 
in Eqs.~(\ref{odv})-\eqref{eq:OEH}, which are made out of two lepton fields and two Higgs doublets. In accordance with the main line of this
section, we do not consider constraints from those operators, which
means that, unless explicitly stated otherwise, when mentioning  $d=6$ operators in this section we refer exclusively to those in
Eqs.~(\ref{equ:ole})-(\ref{equ:oee}).

\subsection{A toy model}

 In order to estimate the theoretical price to pay for obtaining large NSI from exotic particles coupling to SM bilinears, without large charged lepton flavour violation, we show here a toy model in a bottom-up fashion, which precisely generates the $d=8$ operator $\onsi$ in Eq.~(\ref{equ:nsi}) and no $d=6$ operator.
Then we will provide a systematic analysis, from which we will recover the toy model as the simplest possibility in a top-down approach.

Consider the following toy Lagrangian for the underlying theory, which adds both a new scalar doublet (${\bf 2}^s_{1/2}$) 
$\Phi$ and a vector doublet (${\bf 2}^v_{-3/2}$) $V_\mu$  to the SM
Lagrangian, with general couplings  to the SM fields $y$, $g$ and $\lambda$'s,
 \begin{align}
\mathscr{L} & = \mathscr{L}_{\text{SM}}  
  - {(y)_{\beta}}^{\gamma}\, (\bar{L}^\beta)^{i} E_\gamma \Phi_{i} 
  - (g)_{\beta \delta} \, (\bar{L}^\beta)^{i}  \gamma^{\rho}
  (E^c)^\delta (V_\rho)_{i}  
  \nn \\ & 
  \quad 
  + \lambda_{{\bf 1}s}(H^\dagger H)(\Phi^\dagger \Phi)
  + \lambda_{{\bf 3}s}(H^\dagger \vec\tau H)(\Phi^\dagger \vec\tau \Phi) 
  \nn \\
  & \quad 
  + \lambda_{{\bf 1}v}(H^\dagger H) (V_\rho^\dagger V^\rho) 
  + \lambda_{{\bf 3}v}(H^\dagger \vec\tau H)(V_\rho^\dagger \vec\tau
  V^\rho) 
  + \text{h.c.} \,+... 
  \label{equ:toy}
\end{align}
where the dots refer to other bosonic interactions not relevant for this work.
After integrating out the intermediate particles, the following $d=6$ effective interactions involving leptons are induced (see Table~\ref{tab:opdim6}): 
\begin{equation}
\delta \mathscr{L}_{\text{eff}}^{d=6}=\frac{(c^{{\bf 2}v})^{\alpha\gamma}_{\beta\delta}}{\Lambda^{2}} \,(\overline{ E^c}_\gamma \gamma^{\rho} L_\alpha)(\bar{L}^\beta \gamma_{\rho} E^{c\,\delta})    +    \frac{(f^{{\bf 2}s})^{\alpha\gamma}_{\beta\delta}}{\Lambda^{2}} \, (\bar{L}^\beta E_\gamma)(\bar{E}^\delta L_\alpha)\,,
\label{equ:Ltoy6}
\end{equation}
where now 
\be
 \frac{(c^{{\bf 2}v})^{\alpha\gamma}_{\beta\delta}}{\Lambda^{2}} 
 = 
  - \frac{({g}^\dagger)^{\gamma \alpha} (g)_{\beta\delta}}{M^2_V} \, , 
 \quad 
 \frac{(f^{{\bf 2}s})^{\alpha\gamma}_{\beta\delta}}{\Lambda^{2}} 
 = 
 \frac{{({y}^\dagger)_{\delta}}^{\alpha} {(y)_{\beta}}^{\gamma}}{M_\Phi^2} \, .
\ee
For simplicity of notation and illustrative purposes we can assume $M_\Phi \simeq M_V \equiv M (= \Lambda)$.
The $d=6$ cancellation conditions on four charged lepton transitions in \equ{cancelsix}, or its equivalent in the mediator picture 
\equ{cond1}, translate into  
\be \label{equ:cond-1-for-Yukawas}
-\,2
({g}^\dagger)^{\gamma \alpha} (g)_{\beta\delta}
+
{({y}^\dagger)_{\delta}}^{\alpha} {(y)_{\beta}}^{\gamma}
=0\,.
\ee
The relevant effective $d=8$ Lagrangian induced reads 
\begin{align}
\delta \mathscr{L}_{\text{eff}}^{d=8}
=& \frac{1}{M^{4}} [ \lambda_{{\bf 1} s}
\left( \bar{L}\,y \,E \right) 
\left( \bar{E}\,y^\dagger \, L \right) 
( H^\dagger H ) 
+ \lambda_{{\bf 3} s}
\left( \bar{L}\,y\, E \right) \vec{\tau}
\left( \bar{E}\,y^\dagger L \right) 
\left(H^{\dagger} \vec{\tau} H \right) 
\nonumber 
\\
+ & \lambda_{{\bf 1} v}
\left( \bar{L} \,g\, \gamma^{\rho} E^{c} \right) 
\left( \overline{E^{c}} \gamma_{\rho}\,g^\dagger L \right)
(H^{\dagger} H) 
+ \lambda_{{\bf 3} v}
\left( \bar{L} \,g\,\gamma^{\rho} E^{c} \right) \vec{\tau}
\left( \overline{E^{c}} \gamma_{\rho}\,g^\dagger L \right)
\left( H^\dagger \vec{\tau} H \right)],
\label{eq:ToyLeff}
\end{align}
where flavour indices have been omitted and each expression in brackets is to be understood as a flavour singlet.
Eq.~(\ref{eq:ToyLeff}) can be rewritten in terms of the operators of the BR basis in Eqs.~(\ref{equ:br1}) and~(\ref{equ:br2}), as
\begin{align}
\delta \mathscr{L}_{\text{eff}}^{d=8}
=&
 -\frac{1}{\Lambda^4}\,(
 \mathcal{C}^{{\bf 1}}_{LEH} 
\mathcal{O}_{LEH}^{{\bf 1}}
\, + \,\mathcal{C}^{{\bf 3}}_{LEH}
 \mathcal{O}_{LEH}^{{\bf 3}} )\,,
\label{equ:toybr}
\end{align}
where  
\begin{equation}
\mathcal{C}^{{\bf 1}}_{LEH} =
 \lambda_{{\bf 1} v}   \,({g}^\dagger)^{\gamma \alpha} (g)_{\beta\delta}
  +  \frac{1}{2} \lambda_{{\bf 1} s}\,
{({y}^\dagger)_{\delta}}^{\alpha} {(y)_{\beta}}^{\gamma}\,,
\end{equation}
\begin{equation}
\mathcal{C}^{{\bf 3}}_{LEH} =\lambda_{{\bf 3} v}   \,({g}^\dagger)^{\gamma \alpha} (g)_{\beta\delta}
  +   \frac{1}{2}\lambda_{{\bf 3} s}\,
{({y}^\dagger)_{\delta}}^{\alpha} {(y)_{\beta}}^{\gamma}\,.
\end{equation}
In order not to produce interactions between four charged leptons, it is necessary 
to satisfy  \equ{canceleight}, \ie, the condition  
$\mathcal{C}_{LEH}^{{\bf 1}}=\mathcal{C}_{LEH}^{{\bf 3}} \neq 0$, so that the effective $d=8$ interaction in \equ{toybr} reduces precisely  to $\onsi$ in   Eq.~(\ref{nsicomb}).
Together with the $d=6$ cancellation condition, \equ{cond-1-for-Yukawas}, it is finally required that 
\be
\lambda_{{\bf 1} s} + \lambda_{{\bf 1}v} =
\lambda_{{\bf 3} s} + \lambda_{{\bf 3} v} \neq 0\,.
\label{equ:lambda}
\ee
As a consequence, the NSI in matter can be substantial for all flavours. 
While source
and detection NSI cannot be created from our toy model, the epsilon matter parameter  reads
\begin{align}
\left|
\epsilon^{m,R}_{\beta \alpha}
\right|
=&
\frac{v^{4}}{2M^{4}}
\left|
 (\lambda_{{\bf 1} s} + \lambda_{{\bf 1} v})
  (g^{\dagger})^{e\alpha} (g)_{\beta e}
\right|\,.
\end{align}
In resume, by adding both an $SU(2)$ doublet scalar and a doublet vector to the SM content, and imposing two relations to their couplings, 
Eqs.~(\ref{equ:cond-1-for-Yukawas}) and (\ref{equ:lambda}), a toy model for viable large NSI has resulted. 
\begin{figure}[t]
\begin{center}
\unitlength=1.0cm
\begin{picture}(8,6)
  \includegraphics{LLLLHH-scalar.1} 
  \hspace*{0.8cm}
  \includegraphics{LLLLHH-vector.1}
\thicklines
\put(-5.2,0.3){$E$}
\put(-5.2,5.1){$L$}
\put(-8.0,0.3){$L$}
\put(-8.0,5.1){$E$}
\put(-5.2,3.0){$H$}
\put(-8.0,3.0){$H$}
\put(-6.1,3.7){${\bf 2}^s_{+1/2}$}
\put(-6.1,1.5){${\bf 2}^s_{+1/2}$}

\put(-0.6,0.3){$E$}
\put(-0.6,5.1){$E$}
\put(-3.4,0.3){$L$}
\put(-3.4,5.1){$L$}
\put(-0.6,3.0){$H$}
\put(-3.4,3.0){$H$}
\put(-1.5,3.7){${\bf 2}^v_{-3/2}$}
\put(-1.5,1.5){${\bf 2}^v_{-3/2}$}
\end{picture}
\end{center}
\mycaption{\label{fig:model1} Dimension eight operator decomposed into dimension four interactions}
\end{figure}
The model interactions are visualized in \figu{model1}, where the first two
effective interactions in Eq.~(\ref{eq:ToyLeff}) correspond to the diagram on the left 
 -- mediated by  ${\bf 2}^s_{1/2}$ --  and the last two interactions to the diagram on the right
   -- mediated by ${\bf 2}^v_{-3/2}$. 
  In fact, other combinations of just one of the first
two operators in Eq.~(\ref{eq:ToyLeff}) together with one of
the last two operators in that equation would have been enough
for the purpose~\footnote{ For instance a combination involving $\lambda_{{\bf 1}s}$ and $\lambda_{{\bf
3}v}$, or alternatively $\lambda_{{\bf 3}s}$ and $\lambda_{{\bf
1}v}$, would be suitable.}.
 As we will demonstrate below, our toy model is the most general possible model involving only  two mediators,  when the exotic particles couple only to SM bilinears.

 We keep dubbing the construction above as ``toy'' because, to begin
 with, the presence of a vector field,  which is not a gauge boson,
 implies that it is non-renormalizable. The toy Lagrangian, \equ{toy},
 can thus only be considered as an effective theory of some larger
 construction, such as for instance models of extra dimensions in which
 the vector doublet could be a component of a higher dimensional gauge
 theory. 
 
 Moreover, its phenomenological analysis is beyond the scope of the
 present work: the constraints from electroweak precision tests need to
 be analyzed for each specific model,  in particular the {\it oblique} corrections~\cite{Peskin:1990zt, Peskin:1991sw,Hagiwara:1994pw} 
  it may induce. The new couplings may also
 have a relevant impact on other flavour changing transitions at the 
 loop level, although considering large values for the quartic couplings $\lambda$ and small values for  the elements of the $g$ and $y$ flavour matrices, it will probably remain phenomenologically safe.

 The toy model demonstrates that it is possible to achieve the desired $d=8$ interactions, without inducing simultaneously $d=6$ ones, by  fixing the coefficients of the new fields in the Lagrangian. It requires ad-hoc cancellations, though, and it is left as an open question for the model builder whether some symmetry can justify them.

\subsection{Systematic analysis}

 In this subsection, a systematic analysis of all possible effective NSI $d=8$ operators is performed. 
The full decomposition of any combination of $d=8$ operators, constructed from combining bilinear combinations of SM fields, leads to a large number of possibilities. 
 We will first consider  the cases which
are conceptually similar to the toy model above, \ie, new fundamental
interactions involving exactly {\it two}  SM fields, which are the SM bilinears according to our earlier definition. 
Then we will discuss new interactions involving only one SM
field. 

\subsubsection{Mediators coupling to SM bilinears}

We summarize these possibilities for the $\bar{L}L\bar{E}E$-type operators in \Tab~\ref{tab:barLE} and for the $\bar{L}L\bar{L}L$-type operators in \Tab~\ref{tab:barLL2}, which are one of the main results of this study. The notation used has been described in Sect.~3. The tables show, from left to right in each row: 
\begin{itemize}
\item an ordinal assigned to each operator,
\item  the operator itself,
\item the value of the operator coefficients of the BR basis needed to reconstruct it,
\item whether the $d=8$ cancellation conditions in \equ{canceleight} are directly fulfilled (``$\mathcal{\onsi}$?''),
\item the required mediators, with those inducing additional $d=6$ interactions of four charged leptons  (\Tab~\ref{tab:opdim6})  highlighted in boldface.
\end{itemize}
Obviously, the number of possible mediators of $d=8$ interactions
is much larger than for the $d=6$ case in \Tab~\ref{tab:opdim6}. 
In particular, fermions are now possible mediators, unlike for $d=6$. We illustrate the operator decomposition for operator \#2 from
\Tab~\ref{tab:barLE}, showing the
corresponding Feynman diagram in \figu{exdecomp}.

Notice that only the minimal mediator content necessary to obtain
each possible $d=8$ operator is shown in Tables~\ref{tab:barLE} and
~\ref{tab:barLL2}. In other words, although there is always a particular
set of exotic particles whose exchange induces at tree-level the $d=8$
operators considered, this set might not be unique. Nevertheless, for each operator, the particle content shown in the tables is contained in all other possible
sets of mediators leading to it.

From both \Tabs~\ref{tab:barLE} and~\ref{tab:barLL2}, and from \Tab~\ref{tab:opdim6}, one can easily read off the following key results for the operators considered :
\begin{itemize}
\item
There is no way to write down a $d=8$ operator without involving a
     mediator (pinpointed in boldface) which also generates $d=6$
     four-lepton interactions. 
\item
In order to build $\onsi$ {\it and} to cancel the dangerous (or all) NSI $d=6$  contributions, at least two new fields are needed.
\end{itemize}
This implies that  fine-tuning -- or hopefully symmetries -- will be required if all $d=6$ NSI are to be cancelled, \equ{cancelall}~\footnote{ Recall that this  condition ensures that, in addition to avoiding lepton flavour violation among four charged fermions,  
  other putatively dangerous $d=6$ couplings are suppressed, such as for instance possible contributions to the very precise measurement of $G_F$ determined from muon decay.  Note as well that, in principle, one could avoid to impose such a strong cancellation condition by assuming very large couplings among the new heavy fields, and very small values for the couplings between those heavy fields and the SM fields which induce $d=6$ operators. However, since the product between these two types of couplings will be present in the $d=8$ operator (as in our toy model), the $d=8$ couplings would be effectively suppressed as well and extreme fine-tuning would be needed.}.
For model building, one may use the tables as follows: in order to
generate a pure $\onsi$-type operator, it is necessary to choose
effective operators such that \equ{canceleight} is fulfilled, \ie,
interactions with four charged leptons are suppressed, and that
\equ{cancelall} is satisfied, \ie, the NSI contributions from $d=6$
operators cancel. 
The two simplest methods to build a model leading to a pure $\onsi$ interaction are:
\begin{enumerate}
\item
To select from the tables those effective operators marked as $\onsi$.
\item
To linearly combine the effective operators in either of the tables to obtain an $\onsi$ structure.
One possibility is to choose any combination of at least two non-$\onsi$ operators which are linearly independent in the BR basis (not considering $\mathcal{O}_{LLH}^{{\bf 333}}$).\footnote{ In short, the linear combination of two vectors involves only one free parameter (aside from the normalization). The  
condition in \equ{canceleight}  amounts then to a linear equation with only one parameter, which can always be solved for. Since the vectors are linearly independent, they cannot cancel each other, which means that there will be non-vanishing NSI. If, on the other hand, one chooses linearly-dependent vectors, there will be no $d=8$ operator at all -- neither $\onsi$, nor the harmful one.}
\end{enumerate}
The necessary mediators can then be directly read off from the tables; as the next step, the $d=6$ cancellation conditions should be translated into relations among the couplings.

Note that, in addition, there might be flavor dependent conditions and other
constraints, which means that our tables can only serve as hints on how to
build the simplest models. For example, one may have to worry about
electroweak precision data, flavour changing neutral currents,
non-unitarity of the PMNS matrix, loop constraints, and chiral anomalies
if exotic fermions are introduced~\footnote{This  concerns for instance
several examples in~\Tab~\ref{tab:barLE}. In general, in order to cancel the chiral anomaly 
new vector-like fermions may be introduced. In the tables  
we just show the smallest number of  mediators which can induce the $d=8$ operators.}. Also, vectorial scalar $SU(2)$ doublets call for a deeper theory when present, as discussed earlier. 

Other such constraints can result from interactions of the
$\bar{E} E \bar{E} E$-type, which we show
in~\Tab~\ref{tab:EEEE}. Although these interactions do not produce NSI,
care is mandatory when one introduces mediators which could induce such
interactions. For example, operator \#36  not only produces NSI, but
will also lead to potential non-unitarity 
(through the operator in Eq.\eqref{odv}) and other  
unwanted $d=6$ effects, 
and the operator \#61 from \Tab~\ref{tab:EEEE} potentially leads to charged lepton flavor violation.
\begin{figure}[t]
\begin{center}
\unitlength=1.0cm
\begin{picture}(8,6)
\thicklines
\put(0.5,2){\line(1,0){1.5}}
\multiput(2,0.5)(0,0.3){5}{\line(0,1){0.2}}
\put(2,2){\line(4,1){2}}
\put(6,2){\line(-4,1){2}}
\multiput(6,0.5)(0,0.3){5}{\line(0,1){0.2}}
\put(6,2){\line(1,0){1.5}}
\multiput(4,2.5)(0,0.5){4}
{\qbezier(0,0)(0.125,0.125)(0,0.25)
\qbezier(0,0.25)(-0.125,0.375)(0,0.5)}
\put(0.5,4.5){\line(1,0){7}}
\put(0,1.85){$E$}
\put(7.6,1.85){$E$}
\put(1.8,0.1){$H$}
\put(5.8,0.1){$H$}
\put(0,4.4){$L$}
\put(7.6,4.4){$L$}
\put(2.8,1.6){${\bf 2}^{L/R}_{-3/2}$}
\put(4.7,1.6){${\bf 2}^{L/R}_{-3/2}$}
\put(4.3,3.2){${\bf 1}^v_0$}
\put(1.5,2){\vector(1,0){0}}
\put(1.5,4.5){\vector(1,0){0}}
\put(7,2){\vector(1,0){0}}
\put(7,4.5){\vector(1,0){0}}
\put(3,2.25){\vector(4,1){0}}
\put(5,2.25){\vector(4,-1){0}}
\put(2,1.2){\vector(0,1){0}}
\put(6,1.0){\vector(0,-1){0}}
\put(4,4.5){\circle*{0.2}}
\put(2,2){\circle*{0.2}}
\put(6,2){\circle*{0.2}}
\put(4,2.5){\circle*{0.2}}
\end{picture}
\end{center}
\mycaption{\label{fig:exdecomp}Example for a fully decomposed operator. The diagram corresponds to \#2 of \Tab~\ref{tab:barLE}. }
\end{figure}
\begin{table}[p]
\begin{center}
\small{
\begin{tabular}{cccccccc}
\hline \hline
\# 
&
Dim. eight operator 
&
$\mathcal{C}_{LEH}^{{\bf 1}}$
&
$\mathcal{C}_{LEH}^{{\bf 3}}$
& 
$\onsi$?
& 
Mediators
\\
\hline 
\multicolumn{6}{l}{\bf{Combination $\boldsymbol{\bar{L}L}$}} \\
1
&
$(\bar{L} \gamma^{\rho} L) 
(\bar{E} \gamma_{\rho} E) 
(H^{\dagger} H)$
&
$ 1 $
&

&

&
${\bf 1}_{0}^{v}$ 
\\
2 &
$(\bar{L} \gamma^{\rho} L) 
(\bar{E} H^{\dagger})
(\gamma_{\rho})
(H E)$
&
$ 1 $
&

&
&
${\bf 1}_{0}^{v} + { 2}_{-3/2}^{\sss{L/R}}$ 
\\
3 &
$(\bar{L} \gamma^{\rho} L) 
(\bar{E} H^T)
(\gamma_{\rho})
(H^{*} E)$
&
$ 1 $
&

&
& 
${\bf 1}_{0}^{v} + { 2}_{-1/2}^{\sss{L/R}}$
\\
4 &
$(\bar{L} \gamma^{\rho} \vec{\tau} L) 
(\bar{E} \gamma_{\rho} E) 
(H^{\dagger} \vec{\tau} H)$
&

&
1
&
 &
${\bf 3}_{0}^{v} 
+
{\bf 1}_{0}^{v}
$
\\
5 &
$(\bar{L} \gamma^{\rho} \vec{\tau} L) 
(\bar{E} H^{\dagger})
(\gamma_{\rho} \vec{\tau})
(H E)$
&

&
1
&
&
${\bf 3}_{0}^{v} + { 2}_{-3/2}^{\sss{L/R}}$ 
\\
6 &
$(\bar{L} \gamma^{\rho} \vec{\tau} L) 
(\bar{E} H^T)
(\gamma_{\rho} \vec{\tau})
(H^{*} E)$
&

&
1
&
 &
${\bf 3}_{0}^{v} + { 2}_{-1/2}^{\sss{L/R}}$
\\
\hline
\multicolumn{6}{l}{\bf{Combination $\boldsymbol{\bar{E}L}$}} 
\\
7
&
$(\bar{L} E) (\bar{E} L) (H^{\dagger} H)$
&
$-1/2$
&
&
&
${\bf 2}^{s}_{+1/2}$
\\
8
&
$(\bar{L} E) (\vec{\tau}) (\bar{E} L) (H^{\dagger} \vec{\tau} H)$
&
&
$-1/2$
&
&
${\bf 2}^{s}_{+1/2}$
\\
9
&
$(\bar{L} H) (H^{\dagger} E) (\bar{E} L)$
&
$-1/4$
&
$-1/4$
&
$\checkmark$
&
${\bf 2}^{s}_{+1/2}$
+
${ 1}^{\sss{R}}_{0}$
+
${ 2}^{\sss{L/R}}_{-1/2}$
\\
10
&
$(\bar{L} \vec{\tau} H) ( H^{\dagger} E) (\vec{\tau}) (\bar{E} L)$
&
$-3/4$
&
$1/4$
&
&
${\bf 2}^{s}_{+1/2}$
+
${ 3}^{\sss{L/R}}_{0}$
+
${ 2}^{\sss{L/R}}_{-1/2}$
\\
11
&
$(\bar{L} {\rm i}\tau^{2} H^{*}) 
(H^T E) 
({\rm i}\tau^{2}) 
(\bar{E} L)$
&
$1/4$
&
$-1/4$
&
&

${\bf 2}^{s}_{+1/2}$
+
${ 1}^{\sss{L/R}}_{-1}$
+
${ 2}^{\sss{L/R}}_{-3/2}$
\\
12
&
$(\bar{L} \vec{\tau} {\rm i}\tau^{2} H^{*}) 
(H^T E) 
({\rm i}\tau^{2} \vec{\tau}) 
(\bar{E} L)$
&
$3/4$
&
$1/4$
&
&

${\bf 2}^{s}_{+1/2}$
+
${ 3}^{\sss{L/R}}_{-1}$
+
${ 2}^{\sss{L/R}}_{-3/2}$
\\
\hline
\multicolumn{6}{l}{\bf{Combination $\boldsymbol{\overline{E^{c}}L}$}} 
\\
13
&
$(\bar{L} \gamma^{\rho} E^{c}) 
(\overline{E^{c}}\gamma_{\rho} L)
(H^{\dagger} H)$
&
$-1$
&
&
&

${\bf 2}^{v}_{-3/2}$
\\
14
&
$(\bar{L} \gamma^{\rho} E^{c}) (\vec{\tau})
(\overline{E^{c}}\gamma_{\rho} L)
(H^{\dagger} \vec{\tau} H )$
&
&
$-1$
&
&

${\bf 2}^{v}_{-3/2}$
\\
15
&
$(\bar{L} H)
(\gamma^{\rho})
(H^{\dagger} E^c)
(\overline{E^{c}} \gamma_{\rho} L)$
&
$-1/2$
&
$-1/2$
&
$\checkmark$
&
${\bf 2}^{v}_{-3/2}$
+
${ 1}^{\sss{R}}_{0}$
+
${ 2}^{\sss{L/R}}_{+3/2}$
\\
16
&
$(\bar{L} \vec{\tau} H)
(\gamma^{\rho})
( H^{\dagger} E^c)( \vec{\tau})
(\overline{E^{c}} \gamma_{\rho} L)$
&
$-3/2$
&
$1/2$
&
&
${\bf 2}^{v}_{-3/2}$
+
${ 3}^{\sss{L/R}}_{0}$
+
${ 2}^{\sss{L/R}}_{+3/2}$
\\
17
&
$(\bar{L} {\rm i} \tau^{2} H^{*})
(\gamma^{\rho})
(H^T E^c)({\rm i} \tau^{2})
(\overline{E^{c}} \gamma_{\rho} L)$
&
$-1/2$
&
$1/2$
&
&
${\bf 2}^{v}_{-3/2}$
+
${ 1}^{\sss{L/R}}_{-1}$
+
${ 2}^{\sss{L/R}}_{+1/2}$
\\
18
&
$(\bar{L} \vec{\tau} {\rm i} \tau^{2} H^{*})
(\gamma^{\rho})
(H^T E^c) ({\rm i} \tau^{2}\vec{\tau})
(\overline{E^{c}} \gamma_{\rho} L)$
&
$-3/2$
&
$-1/2$
&
&
${\bf 2}^{v}_{-3/2}$
+
${ 3}^{\sss{L/R}}_{-1}$
+
${ 2}^{\sss{L/R}}_{+1/2}$
\\
\hline
\multicolumn{6}{l}{\bf{Combination $\boldsymbol{H^{\dagger} L}$}} 
\\
19
&
$(\bar{L} E)
(\bar{E} H)
(H^{\dagger} L)$
&
$-1/4$
&
$-1/4$
&
$\checkmark$
&
${\bf 2}^{s}_{+1/2}$
+
${ 1}^{\sss{R}}_{0}$
+
${ 2}^{\sss{L/R}}_{-1/2}$
\\
20
&
$(\bar{L} E)
(\vec{\tau})
(\bar{E} H)
(H^{\dagger} \vec{\tau} L)$
&
$-3/4$
&
$1/4$
&
&
${\bf 2}^{s}_{+1/2}$
+
${ 3}^{\sss{L/R}}_{0}$
+
${ 2}^{\sss{L/R}}_{-1/2}$
\\
21
&
$(\bar{L} H)
(\gamma^{\rho})
(H^{\dagger} L)
(\bar{E} \gamma_{\rho} E)$
&
$1/2$
&
$1/2$
&
$\checkmark$
&
${\bf 1}^{v}_{0}$
+
${ 1}^{\sss{R}}_{0}$
\\
22
&
$(\bar{L} \vec{\tau} H)
(\gamma^{\rho})
(H^{\dagger} \vec{\tau} L)
(\bar{E} \gamma_{\rho} E)$
&
$3/2$
&
$-1/2$
&
&
${\bf 1}^{v}_{0}$
+
${ 3}^{\sss{L/R}}_{0}$
\\
23
&
$(\bar{L} \gamma^{\rho} E^{c})
(\overline{E^{c}} H)
(\gamma^{\rho})
(H^{\dagger} L)$
&
$-1/2$
&
$-1/2$
&
$\checkmark$
&
${\bf 2}^{v}_{-3/2}$
+
${ 1}^{\sss{R}}_{0}$
+
${ 2}^{\sss{L/R}}_{+3/2}$
\\
24
&
$(\bar{L} \gamma^{\rho} E^{c})
(\overline{E^{c}} H)
(\gamma^{\rho})
(H^{\dagger} L)$
&
$-3/2$
&
$1/2$
&
&
${\bf 2}^{v}_{-3/2}$
+
${ 3}^{\sss{L/R}}_{0}$
+
${ 2}^{\sss{L/R}}_{+3/2}$
\\
\hline
\multicolumn{6}{l}{\bf{Combination $\boldsymbol{H L}$}} 
\\
25
&
$(\bar{L} E)
({\rm i} \tau^{2})
(\bar{E} H^{*})
(H^T {\rm i} \tau^{2} L)$
&
$1/4$
&
$-1/4$
&
&
${\bf 2}^{s}_{+1/2}$
+
${ 1}^{\sss{L/R}}_{-1}$
+
${ 2}^{\sss{L/R}}_{-3/2}$
\\
26
&
$(\bar{L} E)
(\vec{\tau} {\rm i} \tau^{2})
(\bar{E} H^{*})
(H^T {\rm i} \tau^{2} \vec{\tau} L)$
&
$3/4$
&
$1/4$
&
&
${\bf 2}^{s}_{+1/2}$
+
${ 3}^{\sss{L/R}}_{-1}$
+
${ 2}^{\sss{L/R}}_{-3/2}$
\\
27
&
$(\bar{L} {\rm i} \tau^{2} H^{*})
(\gamma^{\rho})
(H^T {\rm i} \tau^{2} L)
(\bar{E} \gamma_{\rho} E)$
&
$-1/2$
&
$1/2$
&
&
${\bf 1}^{v}_{0}$
+
${ 1}^{\sss{L/R}}_{-1}$
\\
28
&
$(\bar{L} \vec{\tau} {\rm i} \tau^{2} H^{*})
(\gamma^{\rho})
(H^T {\rm i} \tau^{2} \vec{\tau}  L)
(\bar{E} \gamma_{\rho} E)$
&
$-3/2$
&
$-1/2$
&
&
${\bf 1}^{v}_{0}$
+
${ 3}^{\sss{L/R}}_{-1}$
\\
29
&
$(\bar{L} \gamma^{\rho} E^{c})({\rm i} \tau^{2})
(\overline{E^{c}} H^{*})
(\gamma_{\rho})
(H^T {\rm i} \tau^{2}  L)$
&
$1/2$
&
$-1/2$
&
&
${\bf 2}^{v}_{-3/2}$
+
${ 1}^{\sss{L/R}}_{-1}$
+
${ 2}^{\sss{L/R}}_{+1/2}$
\\
30
&
$(\bar{L} \gamma^{\rho} E^{c})( \vec{\tau}{\rm i} \tau^{2})
(\overline{E^{c}} H^{*})
(\gamma_{\rho})
(H^T {\rm i} \tau^{2} \vec{\tau}  L)$
&
$3/2$
&
$1/2$
&
&
${\bf 2}^{v}_{-3/2}$
+
${ 3}^{\sss{L/R}}_{-1}$
+
${ 2}^{\sss{L/R}}_{+1/2}$
\\
\hline 
\hline
\end{tabular}
} 
\end{center}
\mycaption{\label{tab:barLE}
Complete list of $\bar{L} L \bar{E} E$-type $d=8$ interactions which
 involve two SM fields at any possible vertex  of interaction (field bilinears within brackets).  
The columns show an ordinal for each operator, the $d=8$  interaction, the corresponding combination of 
interactions in the BR basis, whether $\onsi$  is satisfied and the
necessary mediators, respectively. Those mediators leading as well to $d=6$ operators in Table 2 are in boldface. The superscript $L/R$ indicates 
massive vector fermions.
The flavor structure  is  to be understood as $\bar{L}^\beta L_\alpha \bar{E}^\delta E_\gamma$.}
\end{table}


\begin{table}[tbp]
\begin{center}
\footnotesize{
\begin{tabular}{cccccccccc}
\hline \hline
\# 
&
Dim. eight operator 
&
$\mathcal{C}_{LLH}^{{\bf 111}}$
&
$\mathcal{C}_{LLH}^{{\bf 331}}$
&
$\mathcal{C}_{LLH}^{{\bf 133}}$
&
$\mathcal{C}_{LLH}^{{\bf 313}}$
&
$\mathcal{C}_{LLH}^{{\bf 333}}$
& 
$\onsi$?
&
Mediators 
\\
\hline 
\multicolumn{6}{l}{\bf{Combination
$\boldsymbol{(\bar{L}^{\beta} L_{\alpha})
(\bar{L}^{\delta} L_{\gamma})(H^{\dagger} H)}$}} \\
31
&
$(\bar{L}\gamma^{\rho}L) 
(\bar{L}\gamma_{\rho}L)
(H^{\dagger} H)$
&
1
&
&
&
&
&
&
${\bf 1}^{v}_{0}$
\\
32
&
$(\bar{L} \gamma^{\rho} \vec{\tau} L) 
(\bar{L} \gamma_{\rho} \vec{\tau} L) 
(H^{\dagger} H)$
&
&
1
&
&
&
&
&
${\bf 3}^{v}_{0}$
\\
33
&
$(\bar{L} \gamma^{\rho} L) 
(\bar{L} \gamma_{\rho} \vec{\tau} L) 
(H^{\dagger} \vec{\tau} H)
$
&
&
&
1
&
&
&
&
${\bf 1}^{v}_{0}$
+
${\bf 3}^{v}_{0}$
\\
34
&
$(\bar{L} \gamma^{\rho} \vec{\tau} L) 
(\bar{L} \gamma_{\rho} L) 
(H^{\dagger} \vec{\tau} H)$
&
&
&
&
1
&
&
&
${\bf 1}^{v}_{0}$
+
${\bf 3}^{v}_{0}$
\\
35
&
$
(-{\rm i} \epsilon^{abc})(\bar{L} \gamma^{\rho} \tau^{a} L ) \times $
&
&
&
&
&
1
&
$\checkmark$
&
${\bf 3}^{v}_{0}$
\\
&
$(\bar{L} \gamma_{\rho} \tau^{b} L ) 
( H^{\dagger} \tau^{c} H )$ \\
\hline 
\multicolumn{6}{l}{\bf{Combination
$\boldsymbol{(\bar{L}^{\beta} L_{\alpha})
(\bar{L}^{\delta} H)(H^{\dagger} L_{\gamma})}$}} 
\\
36
&
$(\bar{L} \gamma^{\rho} L) 
(\bar{L} H) (\gamma_{\rho}) 
(H^{\dagger} L)$
&
1/2
&
&
1/2
&
&
&
$\checkmark$
&
${\bf 1}^{v}_{0}$
+
${1}_{0}^{\sss{R}}$
\\
37
&
$(\bar{L} \gamma^{\rho} L) 
(\bar{L} \vec{\tau} H) 
(\gamma_{\rho}) 
(H^{\dagger} \vec{\tau} L)$
&
3/2
&
&
$-1/2$
&
&
&
&
${\bf 1}^{v}_{0}$
+
${ 3}_{0}^{\sss{L/R}}$
\\
38
&
$(\bar{L} \gamma^{\rho} \vec{\tau} L) 
(\bar{L} \vec{\tau} H)
(\gamma_{\rho}) 
(H^{\dagger} L)$
&
&
$1/2$
&
&
$1/2$
&
$1/2$
&
$\checkmark$
&
${\bf 1}^{v}_{0}$
+
${ 1}_{0}^{\sss{R}}$
+
${ 3}_{0}^{\sss{L/R}}$
\\
39
&
$(\bar{L} \gamma^{\rho} \vec{\tau} L) 
(\bar{L} H) 
(\gamma_{\rho}) 
(H^{\dagger} \vec{\tau} L)$
&
&
$1/2$
&
&
$1/2$
&
$-1/2$
& 
$\checkmark$
&
${\bf 1}^{v}_{0}$
+
${1}_{0}^{\sss{R}}$
+
${3}_{0}^{\sss{L/R}}$
\\
40
&
$(-{\rm i} \epsilon^{abc}) (\bar{L} \gamma^{\rho} \tau^{a} L) \times$
&
&
$1$
&
&
$-1$
&
&
&
${\bf 3}^{v}_{0}$
+
${ 1}_{0}^{\sss{R}}$
+
${ 3}_{0}^{\sss{L/R}}$
\\
& $(\bar{L} \tau^{b} H) 
 ( \gamma_{\rho}) 
(H^{\dagger} \tau^{c} L)$ \\
\hline 
\multicolumn{6}{l}{\bf{Combination
$\boldsymbol{(\bar{L}^{\beta} L_{\alpha})
(\bar{L}^{\delta} H^{\dagger}) (L_{\gamma} H)}$}} 
\\
41
&
$(\bar{L} \gamma^{\rho} L) 
(\bar{L} {\rm i} \tau^{2} H^{*}) 
(\gamma_{\rho})
(H^T {\rm i} \tau^{2} L)$
&
$-1/2$
&
&
$1/2$
&
&
&
&
${\bf 1}^{v}_{0}$
+
${ 1}_{-1}^{\sss{L/R}}$
\\
42
&
$(\bar{L} \gamma^{\rho} L) 
(\bar{L} \vec{\tau} {\rm i} \tau^{2} H^{*}) 
(\gamma_{\rho})
(H^T {\rm i} \tau^{2} \vec{\tau} L)$
&
$-3/2$
&
&
$-1/2$
&
&
&
&
${\bf 1}^{v}_{0}$
+
${3}_{-1}^{\sss{L/R}}$
\\
43
&
$(\bar{L} \gamma^{\rho} \vec{\tau} L) 
(\bar{L} \vec{\tau} {\rm i} \tau^{2} H^{*}) 
(\gamma_{\rho})
(H^T {\rm i} \tau^{2} L)$
&
&
$-1/2$
&
&
$1/2$
&
$1/2$
&
&
${\bf 3}^{v}_{0}$
+
${ 1}_{-1}^{\sss{L/R}}$
+
${ 3}_{-1}^{\sss{L/R}}$
\\
44
&
$(\bar{L} \gamma^{\rho} \vec{\tau} L) 
(\bar{L} {\rm i} \tau^{2} H^{*}) 
(\gamma_{\rho})
(H^T {\rm i} \tau^{2} \vec{\tau} L)$
&
&
$-1/2$
&
&
$1/2$
&
$-1/2$
&
&
${\bf 3}^{v}_{0}$
+
${ 1}_{-1}^{\sss{L/R}}$
+
${ 3}_{-1}^{\sss{L/R}}$
\\
45
&
$(-{\rm i} \epsilon^{abc}) (\bar{L} \gamma^{\rho} \tau^{a} L) \times$
&
&
$-1$
&
&
$-1$
&
&
$\checkmark$
&
${\bf 3}^{v}_{0}$
+
${3}_{-1}^{\sss{L/R}}$
\\
& $ (\bar{L} \tau^{b} {\rm i} \tau^{2} H^{*}) 
( \gamma_{\rho})
(H^T {\rm i} \tau^{2} \tau^{c} L)$
\\
\hline 
\multicolumn{6}{l}{\bf{Combination
$\boldsymbol{
(\bar{L}^{\beta} (L^c)^\delta)
((\overline{L^{c}})_{\alpha} L_{\gamma}) 
(H^{\dagger} H)}$}} 
\\
46
&
$(\bar{L} {\rm i} \tau^{2} L^{c})
(\overline{L^{c}} {\rm i} \tau^{2} L)
(H^{\dagger}H)$
&
$1/4$
&
$-1/4$
&
&
&
&
$\checkmark$
&
${\bf 1}^{s}_{-1}$
\\
47
&
$(\bar{L} \vec{\tau} {\rm i} \tau^{2} L^{c})
(\overline{L^{c}} {\rm i} \tau^{2} \vec{\tau}  L)
(H^{\dagger}H)$
&
$-3/4$
&
$-1/4$
&
&
&
&
&
${\bf 3}^{s}_{-1}$
\\
48
&
$(\bar{L} {\rm i} \tau^{2} L^{c})
(\overline{L^{c}} {\rm i} \tau^{2} \vec{\tau}  L)
(H^{\dagger} \vec{\tau} H)$
&
&
&
$1/4$
&
$-1/4$
&
$-1/4$
&
$\checkmark$
&
${\bf 1}^{s}_{-1}$
+
${\bf 3}^{s}_{-1}$
\\
49
&
$(\bar{L} \vec{\tau} {\rm i} \tau^{2} L^{c})
(\overline{L^{c}} {\rm i} \tau^{2} L)
(H^{\dagger} \vec{\tau} H)$
&
&
&
$-1/4$
&
$1/4$
&
$-1/4$
&
$\checkmark$
&
${\bf 1}^{s}_{-1}$
+
${\bf 3}^{s}_{-1}$
\\
50
&
$ (-{\rm i} \epsilon^{abc})(\bar{L} \tau^{a} {\rm i} \tau^{2} L^{c}) \times$
&
&
&
$-1/2$
&
$-1/2$
&
&
&
${\bf 3}^{s}_{-1}$
\\
& $(\overline{L^{c}} {\rm i} \tau^{2} \tau^{b} L)
( H^{\dagger} \tau^{c} H)$ \\
\hline 
\multicolumn{6}{l}{\bf{Combination
$\boldsymbol{
(\bar{L}^\beta H^{\dagger})
((L^c)^{\delta} H) 
((\overline{L^{c}})_{\alpha} L_{\gamma})}$}} 
\\
51
&
$(\bar{L} {\rm i} \tau^{2} H^{*})
(H^T L^c)
(\overline{L^{c}} {\rm i} \tau^{2} L)$
&
$1/8$
&
$-1/8$
&
$1/8$
&
$-1/8$
&
$1/8$
&
$\checkmark$
&
${\bf 1}^{s}_{-1}$
+
${ 1}^{\sss{L}}_{0}$
+
${ 1}^{\sss{L/R}}_{-1}$
\\
52
&
$(\bar{L} \vec{\tau} {\rm i} \tau^{2} H^{*})
(H^T L^c \vec{\tau})
(\overline{L^{c}} {\rm i} \tau^{2} L)
$
&
$-3/8$
&
$3/8$
&
$1/8$
&
$-1/8$
&
$1/8$
&
$\checkmark$
&
${\bf 1}^{s}_{-1}$
+
${3}^{\sss{L/R}}_{0}$
+
${1}^{\sss{L/R}}_{-1}$
\\
53
&
$(\bar{L} \vec{\tau} {\rm i} \tau^{2} H^{*})
(H^T L^c)
(\overline{L^{c}} {\rm i} \tau^{2}\vec{\tau} L)$
&
$-3/8$
&
$-1/8$
&
$-3/8$
&
$-1/8$
&
$1/8$
&
$\checkmark$
&
${\bf 3}^{s}_{-1}$
+
${ 1}^{\sss{L}}_{0}$
+
${ 3}^{\sss{L/R}}_{-1}$
\\
54
&
$(\bar{L} {\rm i} \tau^{2} H^{*})
(H^T  \vec{\tau} L^c)
(\overline{L^{c}} {\rm i} \tau^{2}\vec{\tau} L)$
&
$3/8$
&
$1/8$
&
$-1/8$
&
$-3/8$
&
$-1/8$
&
&
${\bf 3}^{s}_{-1}$
+
${ 3}^{\sss{L/R}}_{0}$
+
${ 1}^{\sss{L/R}}_{-1}$
\\
55
&
$(-{\rm i} \epsilon^{abc})
(\bar{L} \tau^{a} {\rm i} \tau^{2} H^{*}) \times$
&
$3/4$
&
$1/4$
&
$-1/4$
&
$1/4$
&
$1/4$
&
&
${\bf 3}^{s}_{-1}$
+
${ 3}^{\sss{L/R}}_{0}$
+
${ 1}^{\sss{L/R}}_{-1}$
\\
& $(H^T \tau^{b} L^c)
(\overline{L^{c}} {\rm i} \tau^{2} \tau^{c} L)$ \\
\hline 
\multicolumn{6}{l}{\bf{Combination
$\boldsymbol{
(\bar{L}^{\beta} ({L}^c)^\delta)
(H^{\dagger} (\overline{L^{c}})_{\alpha}) 
(L_{\gamma} H)}$}} 
\\
56
&
$(\bar{L} {\rm i} \tau^{2} L^{c})
(\overline{L^{c}}H^{*})
(H^T {\rm i} \tau^{2} L)$
&
$1/8$
&
$-1/8$
&
$-1/8$
&
$1/8$
&
$1/8$
&
$\checkmark$
&
${\bf 1}^{s}_{-1}$
+
${ 1}^{\sss{L}}_{0}$
+
${ 1}^{\sss{L/R}}_{-1}$
\\
57
&
$(\bar{L} \vec{\tau} {\rm i} \tau^{2} L^{c})
 (\overline{L^{c}}\vec{\tau} H^*)
 (H^T {\rm i} \tau^{2} L)$
&
$3/8$
&
$1/8$
&
$-3/8$
&
$-1/8$
&
$-1/8$
&
&
${\bf 3}^{s}_{-1}$
+
${ 3}^{\sss{L/R}}_{0}$
+
${ 1}^{\sss{L/R}}_{-1}$
\\
58
&
$(\bar{L} {\rm i} \tau^{2} L^{c})
 (\overline{L^{c}} \vec{\tau} H^*)
 (H^T {\rm i} \tau^{2} \vec{\tau}  L)$
&
$-3/8$
&
$3/8$
&
$-1/8$
&
$1/8$
&
$1/8$
&
$\checkmark$
&
${\bf 1}^{s}_{-1}$
+
${ 3}^{\sss{L/R}}_{0}$
+
${ 3}^{\sss{L/R}}_{-1}$
\\
59
&
$(\bar{L} \vec{\tau} {\rm i} \tau^{2} L^{c})
 (\overline{L^{c}} H^*)
 (H^T {\rm i} \tau^{2} \vec{\tau}  L)$
&
$-3/8$
&
$-1/8$
&
$-1/8$
&
$-3/8$
&
$1/8$
&
$\checkmark$
&
${\bf 3}^{s}_{-1}$
+
${ 1}^{\sss{L}}_{0}$
+
${ 3}^{\sss{L/R}}_{-1}$
\\
60
&
$(-{\rm i} \epsilon^{abc})
(\bar{L} \tau^{a} {\rm i} \tau^{2} L^{c}) \times$
&
$3/4$
&
$1/4$
&
$1/4$
&
$-1/4$
&
$1/4$
&
&
${\bf 3}^{s}_{-1}$
+
${ 3}^{\sss{L/R}}_{0}$
+
${ 3}^{\sss{L/R}}_{-1}$ \\
& $( \overline{L^{c}} \tau^{b} H^*)
(H^T {\rm i} \tau^{2} \tau^{c}  L)$
\\
\hline \hline
\end{tabular}
}
\end{center}
\mycaption{\label{tab:barLL2} Same as \Tab~\ref{tab:barLE}, but for the $\bar{L}L\bar{L}L$-type operators.
Note that in this case the relationship between flavor structure and symbol is not unique. We show the
flavor structure for each group separately.}
\end{table}

\begin{table}[tb]
\begin{center}
\small{
\begin{tabular}{cccc}
\hline \hline
\# 
&
Dim. eight operator 
&
$\mathcal{C}_{EEH}$
&
Mediators 
\\
\hline
61
&
$(\bar{E}^{\beta} \gamma^{\rho} E_{\alpha})
(\bar{E}^{\delta} \gamma_{\rho} E_{\gamma}) 
(H^{\dagger} H)$
&
1
&
${\bf 1}_{0}^{v}$
\\
62
&
$(\bar{E} \gamma^{\rho} E)
(\bar{E} H^T)
(\gamma_{\rho}) 
(H^{*} E)$
&
1
&
${\bf 1}_{0}^{v}$
+
${\bf 2}^{\sss{L/R}}_{-1/2}$
\\
63
&
$(\bar{E} \gamma^{\rho} E)
(\bar{E} H^{\dagger})
(\gamma_{\rho}) 
(H E)$
&
$1$
&
${\bf 1}_{0}^{v}$
+
${\bf 2}^{\sss{L/R}}_{-3/2}$
\\
\hline
64
&
$(\bar{E}^{\beta} E^{c \delta})
(\overline{E^{c}}_{\alpha} E_{\gamma})
(H^{\dagger} H)$
&
$1/2$
&
${\bf 1}^{s}_{-2}$
\\
65
&
$(\bar{E} H^{\dagger})
(E^{c} H)
(\overline{E^{c}} E)$
&
$1/2$
&
${\bf 1}^{s}_{-2}$
+
${\bf 2}^{\sss{L/R}}_{-3/2}$
+
${\bf 2}^{\sss{L/R}}_{+1/2}$
\\
66
&
$(\bar{E} E^{c})
(H^{\dagger} \overline{E^{c}})
(E H)$
&
$1/2$
&
${\bf 1}^{s}_{-2}$
+
${\bf 2}^{\sss{L/R}}_{-3/2}$
+
${\bf 2}^{\sss{L/R}}_{+1/2}$
\\
\hline \hline
\end{tabular}
}
\end{center}
\mycaption{\label{tab:EEEE}  Effective $d=8$ operators of the $\bar{E} E \bar{E} E$-type.
The columns shows an ordinal for each operator, the corresponding coefficient in the BR basis and the tree level mediators, respectively. 
The flavour structure is given in the first and fourth rows. Although these operators are not NSI operators, they share with the latter some common mediators which may induce charged lepton flavor violation.}
\end{table}

For the operators in \Tab~\ref{tab:barLE},  our toy model is seen to be 
the only possibility using only two new fields, namely ${\bf
2}^s_{+1/2}$ and ${\bf 2}^v_{-3/2}$. It combines operators \#7, \#8,
\#13, and \#14, which correspond to the four effective interactions in
Eq.~(\ref{eq:ToyLeff}) in our toy model. The table also allows to conclude that it is as well the
most general version of the model with only two fields, while a simpler
version might, for instance, only include \#7 and \#14. Recall that source and 
detection NSI cannot be created from our toy model, while matter NSI for all flavours are allowed. All $\onsi$ operators obtained in Table 3 correspond to the combination of operators of the BR basis in Eq.~(\ref{nsicomb}) which are thus equivalent to Eq.~(\ref{equ:nsi})

In \Tab~\ref{tab:barLL2}, the simplest possibility to build a pure $\onsi$ and no $d=6$ interaction requires
at least three fields, namely ${\bf 1}^s_{-1}$, ${\bf 3}^v_0$, and ${\bf
3}^s_{-1}$, which may come from a large number of possible operator
combinations. For example, one may combine operators \#35 and \#48. As
discussed in \Sec~\ref{formalism}, such a model could have correlations
between source and matter NSI. Note that neither these models, nor our
toy model, involve fermions~\footnote{ More precisely, they do not involve Yukawa couplings linking the exotic and standard fermions.}, which means that they cannot generate
corrections to the unitarity of the PMNS matrix (through contributions to the operators in Eqs.~\eqref{odv} and \eqref{oDv}) 
nor to electroweak data
(through contributions to the operators in Eqs.~\eqref{oDv} and \eqref{eq:OEH}), or at least not at leading order.

\subsubsection{ New interactions involving only one SM field}
\label{Sec:Topology1}

Beyond the operators in the tables above, a much larger number of effective
operators is obtained if, in addition to the interactions with SM bilinears, couplings between one SM field and two
exotic fields are  allowed in the fundamental theory~\cite{Antusch:2008tz}.
The resulting $d=8$ operators are diagrammatically illustrated in
Fig. 4 and fall in three categories,  which contain  the following SM
bilinears at the external vertices:


\begin{figure}[tb]
\unitlength=1.7cm
\begin{picture}(3,3)
\thicklines
\put(1.4,0){\#1}
\put(0.5,1.5){\line(-1,-3){0.4}}
\put(0.5,1.5){\line(-1,3){0.38}}
\put(2.5,1.5){\line(1,-3){0.4}}
\put(2.5,1.5){\line(1,3){0.38}}
\multiput(0.5,1.5)(0.2,0){10}{\line(1,0){0.13}}
\multiput(1,1.5)(0,0.2){6}{\line(0,1){0.13}}
\multiput(2,1.5)(0,0.2){6}{\line(0,1){0.13}}
\put(0.5,1.5){\circle*{0.15}}
\put(2.5,1.5){\circle*{0.15}}
\put(1,1.5){\circle*{0.1}}
\put(2,1.5){\circle*{0.1}}
\put(0,0.1){\footnotesize $L_{\alpha}$}
\put(2.8,0.1){\footnotesize $L_{\beta}$}
\put(0.8,2.7){\footnotesize $H$}
\put(1.8,2.7){\footnotesize $H$}
\put(0.,2.7){\footnotesize $E_{\delta}$}
\put(2.8,2.7){\footnotesize $E_{\gamma}$}
\put(0.33,1){\vector(1,3){0}}
\put(0.305,2.1){\vector(-1,3){0}}
\put(2.74,0.8){\vector(1,-3){0}}
\put(2.665,2){\vector(-1,-3){0}}
\put(1,2.1){\vector(0,1){0}}
\put(2,1.9){\vector(0,-1){0}}
\put(0.85,1.5){\vector(1,0){0}}
\put(2.3,1.5){\vector(1,0){0}}
\put(1.45,1.5){\vector(-1,0){0}}
\put(0.6,1.1){\footnotesize $2^{s}_{1/2}$}
\put(2,1.1){\footnotesize $2^{s}_{1/2}$}
\put(1.4,1.1){\footnotesize $1^{s}_{-1}$}
\end{picture}
\hspace{0.3cm}
\unitlength=1.7cm
\begin{picture}(3,3)
\thicklines
\put(1.4,0){\#2}
\put(0.5,1.5){\line(-1,-3){0.4}}
\put(0.15,1.5){\rotatebox{18}{\multiput(0,0)(0,0.2){6}{\line(0,1){0.13}}}}
\put(2.5,1.5){\line(1,-3){0.4}}
\put(2.5,1.5){\rotatebox{-18}{\multiput(0,0)(0,0.2){6}{\line(0,1){0.13}}}}
\put(0.5,1.5){\line(1,0){0.5}}
\put(2,1.5){\line(1,0){0.5}}
\multiput(1.0,1.5)(0.2,0){5}{\line(1,0){0.13}}
\put(1,1.5){\line(0,1){1}}
\put(2,1.5){\line(0,1){1}}
\put(0.5,1.5){\circle*{0.15}}
\put(2.5,1.5){\circle*{0.15}}
\put(1,1.5){\circle*{0.1}}
\put(2,1.5){\circle*{0.1}}
\put(0.,0.1){\footnotesize $L_{\alpha}$}
\put(2.8,0.1){\footnotesize $L_{\beta}$}
\put(0.8,2.7){\footnotesize $E_{\delta}$}
\put(1.8,2.7){\footnotesize $E_{\gamma}$}
\put(0.,2.7){\footnotesize $H$}
\put(2.8,2.7){\footnotesize $H$}
\put(0.33,1){\vector(1,3){0}}
\put(0.305,2.1){\vector(-1,3){0}}
\put(2.74,0.8){\vector(1,-3){0}}
\put(2.665,2){\vector(-1,-3){0}}
\put(1,2.1){\vector(0,1){0}}
\put(2,1.9){\vector(0,-1){0}}
\put(0.85,1.5){\vector(1,0){0}}
\put(2.3,1.5){\vector(1,0){0}}
\put(1.45,1.5){\vector(-1,0){0}}
\put(0.6,1.2){\footnotesize  $1^R_0$}
\put(2.1,1.2){\footnotesize  $ 1^R_0 $}
\put(1.3,1.2){\footnotesize $ 1^s_{-1}$}
\end{picture}
\hspace{0.3cm}
\unitlength=1.7cm
\begin{picture}(3,3)
\thicklines
\put(1.4,0){\#3}
\put(0.5,1.5){\line(-1,-3){0.4}}
\put(0.15,1.5){\rotatebox{18}{\multiput(0,0)(0,0.2){6}{\line(0,1){0.13}}}}
\put(2.5,1.5){\line(1,-3){0.4}}
\put(2.5,1.5){\rotatebox{-18}{\multiput(0,0)(0,0.2){6}{\line(0,1){0.13}}}}
\put(0.5,1.5){\line(1,0){0.5}}
\put(2,1.5){\line(1,0){0.5}}
\multiput(1.0,1.5)(0.2,0){5}{\line(1,0){0.13}}
\put(1,1.5){\line(0,1){1}}
\put(2,1.5){\line(0,1){1}}
\put(0.5,1.5){\circle*{0.15}}
\put(2.5,1.5){\circle*{0.15}}
\put(1,1.5){\circle*{0.1}}
\put(2,1.5){\circle*{0.1}}
\put(0,0.1){\footnotesize $E_{\gamma}$}
\put(2.8,0.1){\footnotesize $E_{\delta}$}
\put(0.8,2.7){\footnotesize $L_{\beta}$}
\put(1.8,2.7){\footnotesize $L_{\alpha}$}
\put(-0.,2.7){\footnotesize $H$}
\put(2.8,2.7){\footnotesize $H$}
\put(0.33,1){\vector(1,3){0}}
\put(0.305,2.1){\vector(1,-3){0}}
\put(2.74,0.8){\vector(1,-3){0}}
\put(2.74,2.3){\vector(1,3){0}} 
\put(1,2.1){\vector(0,1){0}}
\put(2,1.9){\vector(0,-1){0}}
\put(0.85,1.5){\vector(1,0){0}}
\put(2.3,1.5){\vector(1,0){0}}
\put(1.53,1.5){\vector(1,0){0}}
\put(0.5,1.2){\footnotesize $2^{\sss{L/R}}_{-3/2}$}
\put(2.0,1.2){\footnotesize $2^{\sss{L/R}}_{-3/2}$}
\put(1.4,1.2){\footnotesize $1^{s}_{-1}$}
\end{picture}
\mycaption{\label{fig:1SMfield} Examples for each category of diagrams which lead to $d=8$ operators and require couplings of the new fields {\it both} to SM bilinears and to only one SM field. }
\end{figure}

\begin{enumerate}
\item
 $(LE)$- or $(LL)$-type interactions with new fields. At least one
     of the mediators will necessarily induce some of the $d=6$
     interactions among four  leptons
     discussed earlier (corresponding to the external vertices in the
     figure), and the couplings will thus be subject to the
     corresponding constraints. The fundamental interactions describing
     the internal vertices, however, may not be related to the
     previously discussed $d=6$ interactions.
\item
$(LH)$-type interactions. In this case, the mediators do not
     necessarily induce any dangerous $d=6$ operator involving four leptons, even if there are some common
     mediators. 
          The connections previously studied linking $d=6$ and $d=8$ operators do not need 
       to hold.   Nevertheless, these type of interactions involve exotic fermions ($SU(2)$ singlets or triplets) and are constrained by
    non-unitary contributions to the PMNS matrix and some of them also by electroweak precision data, 
    see, \eg, \Ref~\cite{Antusch:2008tz}:  \figu{1SMfield}, center, illustrates that
     this class of diagrams  is connected to one of the $d=6$ operators in Eqs.\eqref{odv} and \eqref{oDv}, or a combination of them. 
    \item
 $(EH)$-type interactions. These type of interactions are suggestive. The mediators may not induce dangerous  
     four-fermion $d=6$ operators. Furthermore, they do not introduce corrections to the PMNS matrix at leading order.
     They involve exotic leptons, however, which are typically strongly
     constrained by electroweak precision tests~\cite{delAguila:2008pw}.  \figu{1SMfield}, right, illustrates that this class of diagrams is connected to the $d=6$
     operators  in Eq.~\eqref{eq:OEH}.
\end{enumerate}

 Possible ``mixed''
diagrams, that is, diagrams involving two different SM bilinear couplings, will combine the corresponding properties. For instance, a model containing both  $(LE)$ and $(LH)$ couplings to exotic mediators will simultaneously induce  some of the $d=6$ operators in Table~2  {\it and} some of the operators in Eqs.~(10)-(12) which induce non-unitarity. 

 It is easy to show that the vertex involving just one SM field
$(L,E$ or $H$)  requires that the two exotic particles attached to it
have different $SU(2)\times U(1)$ charges.
Indeed, we have explicitly checked that {\it all} of these possibilities require at least two new
fields to be phenomenologically viable, \ie, are not simpler than the cases discussed prior to this subsection.

  The scenarios in diagram \#2 and specially  \#3 in \figu{1SMfield} are appealing alternatives, as none of them is correlated to harmful $d=6$ interactions (\ie, four charged-fermion lepton couplings), and \#3 does not induce non-unitarity either. Furthermore, these two examples are $\onsi$  operators. Indeed, the exchange of a singlet fermion ${\bf 1}_0^R$ and a charged scalar ${\bf 1}_{-1}^s$ shown in \#2  gives schematically
  \begin{equation}
  (\bar{L} H)  (E) (\bar{E}) (H^{\dagger} L) =  -\frac{1}{4}
 (\mathcal{O}_{LEH}^{\bf 1})
 -\frac{1}{4}
 (\mathcal{O}_{LEH}^{\bf 3})\,.
  \end{equation}
  Here the projection onto the BR basis shows that it  complies with the $d=8$ cancellation conditions, \equ{canceleight}. The mediator ${\bf 1}_{-1}^s$ could induce in addition $d=6$ effective interactions if it would also couple to SM lepton doublets, as shown in \Tab~\ref{tab:opdim6}, but such couplings are not mandatory. In contrast, the PMNS unitarity constraints should be relevant, as a singlet exotic fermion is involved.
  
   Turning now to type \#3 and the scenario with an exotic doublet fermion ${\bf 2}^{\sss{L/R}}_{-3/2}$ and a charged scalar ${\bf 1}_{-1}^s$, the resulting effective operator for this example is  of the form
   \begin{equation}
   (E H) 
 (\bar{L})
 (L)
 (H^{\dagger} \bar{E})= -\frac{1}{4}
 (\mathcal{O}_{LEH}^{\bf 1})
 -\frac{1}{4}
 (\mathcal{O}_{LEH}^{\bf 3})\,,
  \end{equation}
 and is thus again of the $\onsi$ type. Furthermore, in this case the interactions neither  lead to non-unitarity, nor any $d=6$ operator in \Tab~\ref{tab:opdim6} needs to be induced if the charged scalar does not couple to SM lepton doublets (in other words, the $d=6$ complete cancellation conditions in \equ{cancelall} can be implemented as well). 
 Other scenarios of the kind just discussed do not necessarily have to lead by themselves to $\onsi$ structures: for them, cancellations similar to those in our toy model could be considered. 
 However, it remains to be explored how difficult is to circumvent the constraints which 
electroweak precision tests impose on exotic leptons, and whether the 
necessary cancellations are feasible without running into  
extreme fine-tunings, for instance enlarging the scalar sector of the theory.

 
During the completion of this work,  \Ref~\cite{Antusch:2008tz} appeared.
It explores (but is not limited to) the possible exchange of exotic
fields which
in our notation have quantum numbers of a scalar ${\bf 1}^s_{-1}$ (to
obtain 
$d=6$ NSI) and  of a fermion ${\bf 1}_0^R$ (to obtain $d=8$ NSI). The
latter
induces also $d=6$ interactions, which lead to non-unitary contributions
to the
PMNS matrix, as it is well known and is further explored in that reference.
\Ref~\cite{Antusch:2008tz} performs a systematic topological scan of the
$d=8$ operators, based on Feynman diagrams, trying to obtain the
interaction $\onsi$ directly  from just one Feynman diagram while avoiding
{\em any} harmful $d=6$ and $d=8$ contribution. Our tables correspond to the
topologies~2
and~3 in this reference, whereas the previous paragraph in this
subsection would
correspond to their topology~1. Since all possibilities in our tables
contain
at least one mediator leading to harmful $d=6$ effects if one does not allow
for cancellations, \Ref~\cite{Antusch:2008tz} effectively exclude
topologies~2
and~3 in their scan (apart from our \#46, which does not induce harmful $d=6$
four charged lepton interactions, but the mediator ${\bf 1}^s_{-1}$ is
constrained otherwise, as we and \Ref~\cite{Antusch:2008tz} discussed before). Therefore, our work
is complementary to that reference. Note that they find that the NSI in
matter and the NSI at source or
 detector are correlated in all of their examples by the non-unitary
effects of the heavy fermions,
whereas it is easy to see that uncorrelated scenarios are achievable
when one allows to
combine different operators from our Table \ref{tab:barLE} (such as \#7,
\#8, \#13 and \#14).
 As the most important difference, we relate the operators obtained from
mediator exchanges to
a complete basis of independent operators, which allows us to deduce the
general
cancellation conditions. 

\section{Summary and conclusion}

In this study, we have discussed the possibility of large non-standard interactions (NSI) in the neutrino sector. Since any model of new physics has to recover the Standard Model at low energies, we have required gauge invariance under the SM gauge group and studied the possible effective theories. The focus is set  on purely leptonic NSI, that is,  on operators in which the only fermion fields appearing are leptons. Our analysis has been based on the full (analytical) decomposition of all possible dimension six and eight effective operators, which can be induced at tree-level by any hypothetical beyond the SM theory. Special focus has been set in the scenario in which the exotic mediators couple to SM bilinear field combinations.

The aim is to gauge the theoretical price of achieving phenomenologically viable large neutrino NSI, and to establish the minimal constraints that models have to respect for this purpose. Our main requirements were:
\begin{itemize}
\item
Interactions with four charged leptons have to be absent or highly suppressed, since these would lead to charged lepton flavor violation or corrections to $G_F$.
\item
When analyzing NSI from $d=8$ operators, any $d=6$ contribution among four leptons is not allowed or has to be very suppressed, since this would either be the dominating NSI (if harmless), or lead to unacceptably strong interactions among four charged leptons (if harmful).
\end{itemize}
The NSI operators obtained have been expanded in a complete basis of
independent operators, which has allowed us to consistently consider
cancellations among the contributions of different operators. This new
approach has established the general cancellation conditions which the
model parameters have to fulfill, to avoid four charged lepton interactions when the exotic mediators couple to SM bilinears.

We have then studied the required complexity of any realistic model, such as what is the number of mediators and/or the type of cancellations needed. In short, we have demonstrated that is not possible to create such NSI at $d=6$ for all flavour channels. For $d=8$, we have constructed a simple toy model in which the necessary cancellations  occur. It introduces two new $SU(2)$ doublet mediators, a Lorentz scalar and a vector, which induce the desired large NSI without any dangerous $d=6$ flavour-changing transitions among  four charged leptons or non-unitarity corrections. It also allows for matter 
NSI uncorrelated with source or detector effects. 
 Furthermore, we have obtained and analyzed the general classification of $d=8$ interactions in a systematic way. 

More precisely, for the $d=6$ operators it is shown at the effective operator level that, 
if the four charged lepton contributions have to be exactly cancelled,
   it is not possible to obtain $\eet$ under the above assumptions. In addition, there are certain connections between the source NSI at a neutrino factory, and the matter NSI, such as $\epsilon^m_{\mu \tau} = - (\epsilon^s_{\mu \tau})^{*}$. We have shown that there is only one viable possibility for a $d=6$ interaction with {\em one mediator only}, which is the well-known antisymmetric operator from \Ref~\cite{Bergmann:1999pk}. Nevertheless, there are other options with more than one mediator in order to cancel  all related interactions involving four charged leptons. 
 Altogether, if the full decomposition of the operators is taken into account together with the current bounds, only $\ett$ might be viable with only one mediator and be as large as order unity. In this case, the current bounds do not require that the four charged lepton contributions cancel exactly.

For the $d=8$ operators, we have shown that at least two new fields are required to avoid the undesired $d=6$ and $d=8$ interactions involving four charged leptons. In fact, when the mediators of a $d=8$ effective operator couple only to SM bilinears,
there will always be at least one field leading as well to $d=6$ contributions. These 
have to be cancelled in each case by fine-tuning or symmetries. This result can be directly seen from \Tab~\ref{tab:barLE} and \Tab~\ref{tab:barLL2}, which list the possible dimension eight operators, including the mediators, for fundamental interactions involving two SM fields. 
 In those tables, the mediators highlighted in boldface lead to dimension six interactions as well. 

Our simple toy model, with two bosonic $SU(2)$ doublets added to the SM content, leads  directly to the desired $d=8$ NSI  and fulfills
our minimal requirements. Notice, however, that we have not considered
constraints from loop effects, neither for this toy model nor for the
general operator analysis, as it is beyond the scope of the present
work. These corrections could be specially relevant for some models when considering $d=8$ interactions, 
as the one-loop corrections could spoil the $d=6$ cancellation conditions.
They should be taken into account in any  model which aims to be realistic.

Similar considerations apply to the scenarios in which exotic couplings to {\it both} one SM field and SM bilinears are simultaneously allowed. In order to induce then large $d=8$ NSI and no $d=6$ couplings among four leptons, a minimum of two exotic mediators is once again needed. Some simple candidate models may not even require (strong) cancellation conditions and deserve further exploration. This is the case, for instance, when a singlet scalar and a fermionic doublet are added to the SM content.

 As far as the connection between source and matter NSI is concerned, we have demonstrated that it depends on the operators used.  
   For example, several of the $d=8$ operators in \Tab~\ref{tab:barLE}, or  combinations of them, will only induce matter NSI, while those requiring singlet or triplet fermionic mediators may induce correlations (through non-unitary corrections to the PMNS matrix). On the other hand, all $d=8$ operators in \Tab~\ref{tab:barLL2} will, in principle, allow for a connection between source and matter NSI independent of the mediators used. Therefore, it might be very well possible to detect matter NSI without source or detector effects, in the absence of fermions as exotic particles, such as illustrated by our toy model. Note as well that the models based on \Tab~\ref{tab:barLL2} require at least three new fields, which means that a source and matter NSI connection might be more easily achieved through non-unitary corrections to the PMNS matrix.

In conclusion, we have demonstrated that the minimum complexity of a realistic model leading to large NSI and no charged lepton flavor violation requires at least two new fields inducing $d=8$ NSI couplings. We have determined the possible SM charges of those mediators and the cancellation conditions for the dimension six  interactions among four leptons that they simultaneously induce in most cases. These cancellation conditions translate into precise relations among model parameters. One exception might be $\ett$, which might be created at the dimension six level. 
 Our results imply a number of constraints such that the 
observational prospects do not seem bright, specially as we did not identify some symmetry which would 
account for them. On the other side, we showed that large NSI are not excluded,
and we found out which conditions are necessary to satisfy for any model to be viable.
We agree that those conditions should be justified by symmetries or other arguments 
for the model to be credible. Until such justification is maybe found in some model, 
we leave it up to the reader to decide on the perspective for large NSI.

\subsection*{Acknowledgments}

T.O. and W.W. would like to thank the theory group at UAM for their hospitality, where a part of this work was carried out.
M.B.G. acknowledges illuminating discussions with Enrique Fern\'andez-Mart\'{\i}nez and Pilar Hern\'andez. 
Furthermore, M.B.G. and D.H. received partial
support from CICYT through the project FPA2006-05423, as well as from
the Comunidad Aut\'onoma de Madrid through Proyecto HEPHACOS; D.H. acknowledges financial support from The MEC through FPU grant AP20053603.  
T.O. and W.W. would also like to acknowledge support from Emmy Noether program of Deutsche Forschungsgemeinschaft.  D. H. acknowledges as well the LPT (Orsay) for hospitality during the last stage of this work.




\begin{appendix}
\section{On non-standard four neutrino interactions} 
\label{app:fournu}

Although interactions among four neutrinos hardly contribute to laboratory processes,
there has been some interest in the literature in the context of flavor 
oscillations in astrophysical environments, such as dense neutrino gases; see \eg\ 
 \Ref~\cite{Blennow:2008er} and references therein.
The direct laboratory bounds on these interactions are naturally extremely weak, see 
\Refs~\cite{Bilenky:1999dn,Belotsky:2001fb}.
In this appendix, we discuss these four neutrino interactions in our gauge invariant
framework.

\subsubsection*{Effective operator formalism}

Since the four neutrino interactions require interactions with four
lepton doublets, they only appear for the $\bar{L} L \bar{L} L$ operators. In this
case, \equ{LLLL} reads, including the four neutrino interactions,
\begin{align}
\delta \mathscr{L}_{\text{eff}}
&= 
 \frac{1}{\Lambda^{2}}
 \left( 
 \mathcal{C}_{\mathrm{NSI}}^{\bar{L}L\bar{L}L} 
 \right)^{\alpha \gamma}_{\beta \delta} 
 \left(
 \bar{\nu}^{\beta} \gamma^{\rho} {\rm P}_{L} \nu_{\alpha}
  \right) \, 
 \left(
 \bar{\ell}^{\delta} \gamma^{\rho} {\rm P}_{L} \ell_{\gamma} 
 \right)
 \nonumber \\
 &+
 \frac{1}{\Lambda^{2}}
 \left(
 \mathcal{C}_{LL}^{{\bf 1}}
 +
 \mathcal{C}_{LL}^{{\bf 3}}
 +
 \frac{v^{2}}{2\Lambda^{2}} \left(
 \mathcal{C}_{LLH}^{{\bf 111}}
 +
 \mathcal{C}_{LLH}^{{\bf 331}}
 -
 \mathcal{C}_{LLH}^{{\bf 133}}
 -
 \mathcal{C}_{LLH}^{{\bf 313}} \right)
 \right)_{\beta \delta}^{\alpha \gamma}
 \left(
 \bar{\ell}^{\beta} \gamma^{\rho} {\rm P}_{L} \ell_{\alpha}
 \right)
 \left(
 \bar{\ell}^{\delta} \gamma^{\rho} {\rm P}_{L} \ell_{\gamma} 
 \right) \nonumber \\
 &+
 \frac{1}{\Lambda^{2}}
 \left(
 \mathcal{C}_{LL}^{{\bf 1}}
 +
 \mathcal{C}_{LL}^{{\bf 3}}
 +
 \frac{v^{2}}{2\Lambda^{2}} \left(
 \mathcal{C}_{LLH}^{{\bf 111}}
 +
 \mathcal{C}_{LLH}^{{\bf 331}}
 +
 \mathcal{C}_{LLH}^{{\bf 133}}
 +
 \mathcal{C}_{LLH}^{{\bf 313}} \right)
 \right)_{\beta \delta}^{\alpha \gamma}
 \left(
 \bar{\nu}^{\beta} \gamma^{\rho} {\rm P}_{L} \nu_{\alpha}
 \right)
 \left(
 \bar{\nu}^{\delta} \gamma^{\rho} {\rm P}_{L} \nu_{\gamma} 
 \right) \nonumber \\
&+ \mathrm{h.c.} \, . \label{equ:LLLL4nu}
\end{align}

The first point one notices is that the four charged lepton and four neutrino interactions
share for $d=6$ the same coefficient $\mathcal{C}_{LL}^{{\bf 1}}+\mathcal{C}_{LL}^{{\bf 3}}$.
This means that for $d=6$, any bound from charged lepton flavor violation \etc\ can be
directly translated into the four neutrino interactions. This is illustrated here with 
one example. For $\beta=\mu$ and $\alpha=\gamma=\delta=e$, the bound from $\mu \rightarrow eee$
can, apart from some SU(2) symmetry breaking effects, be directly transferred to the four 
neutrino interactions. In our notation, one has
\begin{equation}
 \mathrm{Br}(\mu \rightarrow 3e) = \frac{1}{G_{F}^{2}} \, \left( \frac{ \mathcal{C}_{LL}^{{\bf 1}}
 +
 \mathcal{C}_{LL}^{{\bf 3}}}{\Lambda^{2}} \right)^{2}  = \frac{F^2}{G_F^2} \, ,
\end{equation}
where the non-standard parameter is defined as $F \equiv (\mathcal{C}_{LL}^{{\bf 1}}
 + \mathcal{C}_{LL}^{{\bf 3}})/\Lambda^{2}$ -- as often done in the literature. The current bound $\mathrm{Br}(\mu \rightarrow 3e) < 10^{-12}$ (90\% CL)~\cite{Amsler:2008zz} then directly translates into $F \lesssim 10^{-6} \, G_F$,
which is far below any laboratory bound or even the bound from primordial nucleosynthesis.  
Of course, it is dependent on the participating flavors and somewhat looser for combinations involving the $\tau$, but this procedure illustrates the generic argument.
Note that the bound for a vector mediated interaction, such as often discussed in the literature, turns out to be the same in this case. 

As discussed in \Sec~\ref{formalism},  \equ{cancelsix} should be satisfied for
any realistic model in order to avoid these bounds. As we can read off from \equ{LLLL4nu},
however, the $d=6$ coefficients for the four charged lepton and four neutrino interactions
are exactly the same, which means that there will not be any four neutrino interactions
in that case.
As a consequence, one has to go to $d=8$ with the interactions
being suppressed by $\Lambda^4$.

For $d=8$, the corresponding \equ{canceleight} to suppress the harmful interactions among four charged fermions
 can be implemented in qualitatively different ways. For example,
if $\mathcal{C}_{LLH}^{{\bf 111}}
 = - \mathcal{C}_{LLH}^{{\bf 331}}$ and $
 \mathcal{C}_{LLH}^{{\bf 133}}
 = -
 \mathcal{C}_{LLH}^{{\bf 313}}$, there will be no four neutrino interactions but NSI,
whereas for 
$\mathcal{C}_{LLH}^{{\bf 111}}
 +
 \mathcal{C}_{LLH}^{{\bf 331}}
 =
 \mathcal{C}_{LLH}^{{\bf 133}}
 +
 \mathcal{C}_{LLH}^{{\bf 313}}  \neq 0$, there will be both four neutrino interactions and NSI. 
As it is demonstrated below, both possibilities can be realized within the model framework in this study.

\subsubsection*{Model analysis}

In order to find models for large four neutrino interactions at $d=8$,
the same argumentation as in \Sec~\ref{sec:eight} is needed. First of all,
\equ{canceleight} has to be satisfied to suppress the four charged
lepton processes. Second, the $d=6$ contributions to the NSI have to be cancelled,
since there are strong bounds, \ie, \equ{cancelall} has to be satisfied.
As an additional condition, one can {\bf not} have (\cf, \equ{LLLL4nu}) 
\begin{equation}
 \mathcal{C}_{LLH}^{{\bf 111}}
 +
 \mathcal{C}_{LLH}^{{\bf 331}}
 +
 \mathcal{C}_{LLH}^{{\bf 133}}
 +
 \mathcal{C}_{LLH}^{{\bf 313}}  = 0 \, 
\label{equ:fournu}
\end{equation}
because such an operator will not contribute to the four neutrino interactions.
The relevant decomposed operators can be found in \Tab~\ref{tab:barLL2}, where
one can easily read off if \equ{fournu} is satisfied. Furthermore,
note that
operators which only induce $\mathcal{C}_{LLH}^{{\bf 333}}$ will not
be useful for the four neutrino interactions. We find from the table that 
operators \#35, \#40, \#41, \#43, \#44, \#46, \#48, \#49, \#51, \#52, \#54,
\#56, \#57, and \#58 do not contribute to the four neutrino interactions. 
This implies that the possibility pointed out in the
main text, \ie, to combine \#35 and \#48, does not lead to
four neutrino interactions. One has to use more complicated
combinations by the combination of different operators, such as \#32 and 
\#50 to satisfy \equ{canceleight}, and \#48 (which satisfies \equ{canceleight}) to 
introduce an additional mediator to cancel the $d=6$ NSI. Then the
four neutrino interactions can be constructed with three different mediators, where
only \#32 and \#50 contribute to the four neutrino interactions. 
Constructions with less mediators are, under the assumptions in this study, not possible,
which is different from the NSI, which can be generated from two mediators. 

As soon as a specific model is known, the relationship among source and production
NSI, matter NSI, and four neutrino interactions can be easily calculated using
\Sec~\ref{formalism} and \equ{LLLL4nu}. 

In summary, for the $d=6$ four neutrino interactions, gauge invariance
implies that they face the stringent bounds from charged lepton flavor violation,
such as from $\mu$ to three electrons. Therefore, large four neutrino interactions
have to come from $d=8$ effective operators. From the model point of view,
having four neutrino interactions is even more complicated than having large
NSI, since at least three different mediators are needed in the framework discussed in this study.

\end{appendix}

\end{document}